\documentclass[a4paper,11pt]{article}
\pdfoutput=1 % if your are submitting a pdflatex (i.e. if you have
\usepackage{jcappub} % for details on the use of the package, please
\usepackage{aasmacros}

\usepackage{natbib}
\usepackage[utf8]{inputenc} % allow utf-8 input
\usepackage{hyperref}       % hyperlinks
\usepackage{url}            % simple URL typesetting
\usepackage[dvipsnames]{xcolor}
\usepackage{subfigure}
\usepackage{caption}
\usepackage{subcaption}
\usepackage{graphicx}
\usepackage{lineno}
\usepackage{multirow}
\usepackage{amsmath}
\usepackage{amssymb}
\usepackage{lipsum}  
\usepackage{verbatim}
\usepackage[normalem]{ulem}

\newcommand{\Msun}{{\rm M}_\odot}

\title{The Large Magellanic Cloud: expanding the low-mass parameter space of dark matter direct detection}

\author[a]{Javier Reynoso-Cordova,}
\author[a,b]{Nassim Bozorgnia,}
\author[a]{and Marie-Cécile Piro}
\affiliation[a]{Department of Physics, University of Alberta,\\
CCIS 4-181, Edmonton, Alberta T6G 2E1, Canada}
\affiliation[b]{Theoretical Physics Institute, University of Alberta,\\
CCIS 4-181, Edmonton, Alberta T6G 2E1, Canada}
\emailAdd{jireynos@ualberta.ca}
\emailAdd{nbozorgnia@ualberta.ca}
\emailAdd{mariecci@ualberta.ca}

\abstract{We investigate how the Large Magellanic Cloud (LMC) impacts the predicted signals in near-future direct detection experiments for non-standard dark matter (DM) interactions, using the Auriga cosmological simulations. We extract the local DM distribution of a simulated Milky Way-like halo that has an LMC analogue and study the expected signals in DarkSide-20k, SBC, DARWIN$\slash$XLZD, SuperCDMS, NEWS-G, and DarkSPHERE considering DM-nucleon effective interactions, as well as inelastic DM scattering. We find that the LMC causes substantial shifts in direct detection exclusion limits towards smaller cross sections and DM masses for all non-relativistic effective field theory (NREFT) operators, with the impact being highly pronounced for velocity-dependent operators at low DM masses. For inelastic DM, where the DM particle up-scatters to a heavier state, the LMC  shifts the direct detection exclusion limits towards larger DM mass splitting and smaller cross sections. Thus, we show that the LMC significantly expands the parameter space that can be probed by direct detection experiments towards smaller DM-nucleon cross sections for all NREFT operators and larger values of mass splitting for inelastic DM.}

\begin{document}
\maketitle
\flushbottom

\section{Introduction}
\label{sec:introduction}

Cosmological and astrophysical observations from various scales have shown that a cold dark matter (DM) component is fundamental to the standard $\Lambda$CDM ($\Lambda$-cold dark matter) cosmological model~\cite{Planck:2018vyg}. Despite all the evidence, we do not yet have an unambiguous signal which sheds light onto the nature or composition of DM. Within the plateau of possible DM candidates, Weakly Interacting Massive Particles (WIMPs) are the most extensively studied due to their implications on the weak sector of the Standard Model of particle physics~\cite{Feng:2010gw}. During the past decades, a considerable effort has been made to detect such a particle.

Direct detection experiments offer the opportunity to measure the signal that a WIMP would produce when scattering with the target nucleus or electron in an underground detector~\cite{Goodman:1984dc}. These experiments utilize different targets to explore various types of interactions  probing DM masses ranging from tens of GeV up to TeV-scale~\cite{Jungman:1995df}. However, the ongoing search for WIMPs has prompted the exploration of parameter spaces with masses lighter than the traditional 10 GeV range~\cite{Drukier:2013xtn}. In particular, noble-liquid~\cite{DarkSide-50:2022qzh, DarkSide-20k:2017zyg, Alfonso-Pita:2023frp, Schumann:2015cpa, XENON:2023cxc, Aalbers:2022dzr, Baudis:2024jnk}, cryogenic~\cite{SuperCDMS:2014cds, SuperCDMS:2015lcz, SuperCDMS:2016wui, SuperCDMS:2017nns}, and gaseous~\cite{NEWS-G:2017pxg, arora2024searchlightdarkmatter, Durnford:2021mzg, NEWS-G:2023qwh} experiments have been established as attractive detector technologies to set the most stringent constraints on the spin-independent and spin-dependent DM-nucleon cross sections for DM direct detection searches.

An important input in the analysis of direct detection results is the DM phase-space distribution in our Solar neighborhood. In the simplest and most commonly adopted halo model, the local DM velocity distribution is assumed to be a Maxwell-Boltzmann distribution with a peak speed set to the local circular speed in the Milky Way (MW). This model is known as the Standard Halo Model (SHM) \cite{PhysRevD.33.3495}. The actual DM distribution, however, can be substantially different from the SHM. High resolution cosmological simulations including both DM and baryons show that although a Maxwellian velocity distribution provides a good fit to the local DM velocity distribution of simulated MW-like halos, there exists large uncertainties in the results~\cite{Bozorgnia:2016ogo, Kelso:2016qqj, Sloane:2016kyi,  Bozorgnia:2017brl,  Bozorgnia:2019mjk, Poole-McKenzie:2020dbo, Lawrence:2022niq, Kuhlen:2013tra, Lacroix:2020lhn}.

One of the most significant changes to the SHM arises from the impact of the Large Magellanic Cloud (LMC) on the local DM distribution. Recent work using both  cosmological~\cite{Smith-Orlik:2023kyl} and idealized~\cite{Besla:2019xbx, Donaldson:2021byu} simulations have shown that the LMC can significantly boost the local DM velocity distribution to speeds greater than the escape speed from the Galaxy. This leads to substantial shifts of several orders of magnitude in the direct detection exclusion limits towards smaller cross sections and DM masses~\cite{Smith-Orlik:2023kyl}, especially for low mass DM. These shifts result from the existence of the DM particles originating from the LMC in the Solar neighborhood, as well as the gravitational potential of the LMC boosting the DM particles of the MW.

Given the major impact of the LMC on DM direct detection limits in the case of the simplest standard spin-independent DM-nucleon interactions, it remains to be examined how the LMC affects non-standard DM interactions. For instance several theoretical models predict the existence of mediators, which translate into a dependence of the DM-nucleon cross section on the momentum transfer or relative velocity between the DM and nucleon in the non-relativistic limit. These non-standard interactions can be described using a non-relativistic effective field theory (NREFT) approach, which  parametrizes the DM-nucleus interaction in terms of a set of operators that appear in the effective Lagrangian describing the interaction~\cite{Fan:2010gt, Fitzpatrick:2012ix, Fitzpatrick:2012ib, Anand:2013yka, Dent:2015zpa}. Some of these operators lead to different dependence of the DM-nucleon cross section on the DM velocity. A modified DM velocity distribution due to the LMC can therefore have a more significant impact on direct detection limits on such DM-nucleon effective couplings. On the other hand, in the case of inelastic DM scattering~\cite{Tucker-Smith:2001myb}, where DM up-scatters to a heavier state, the experiments become more sensitive to the high speed tail of the local DM velocity distribution. Consequently, the LMC's impact on direct detection limits would be more pronounced in this case.

In this paper, we use the Auriga cosmological magneto-hydrodynamical simulations~\cite{Grand:2016mgo} to study the impact of the LMC on the interpretation of results from the next-generation direct detection experiments in the context of DM-nucleon effective interactions as well as DM inelastic scattering. In particular, we focus on six near-future experiments with different target nuclei to cover a large range of atomic mass number, $A$, and compare the relative changes in different experiments: DarkSide-20k~\cite{DarkSide-20k:2017zyg} and SBC (Scintillating Bubble Chamber)~\cite{Alfonso-Pita:2023frp} with argon ($A = 40$), DARWIN$\slash$XLZD~\cite{Schumann:2015cpa, Aalbers:2022dzr, Baudis:2024jnk} with xenon ($A = 131$), SuperCDMS~\cite{SuperCDMS:2016wui} with germanium ($A = 72$), NEWS-G (New Experiments With Spheres-Gas)~\cite{NEWS-G:2017pxg, Durnford:2021mzg} with neon ($A = 20$), and DarkSPHERE \cite{NEWS-G:2023qwh} with helium ($A = 4$) target nuclei. The paper is structured as follows. In section~\ref{sec:sims} we discuss the details of the simulations and the MW-LMC analogue that we use. In section~\ref{sec:fv} we present the local DM distribution of the MW analogue. In section~\ref{sec:directdetection} we discuss the calculations of DM direct detection signals, with the halo integrals given in section~\ref{sec:halo} and the NREFT operators discussed in section~\ref{sec:NREFT}. In section~\ref{sec:experiments} we discuss the parameters of the six near-future experiments we consider. In section~\ref{sec:results} we present our results, and we conclude in section~\ref{sec:conclusions}.

\section{Simulations}
\label{sec:sims}

In this work, we use the simulated MW-LMC system first studied in ref.~\cite{Smith-Orlik:2023kyl}. This system is identified from the Auriga magneto-hydrodynamical simulations~\citep{Grand:2016mgo, Grand:2024}. The Auriga project~\citep{Grand:2016mgo} includes a set of zoom-in simulations of isolated MW mass halos, which were selected from a $100^3$~Mpc$^3$ periodic cube (L100N1504) from the EAGLE project~\cite{Schaye:2014tpa, Crain:2015poa}. The simulations were performed using the moving-mesh code Arepo~\citep{Springel:2009aa} and implement a galaxy formation subgrid model which includes active galactic nuclei and supernova feedback, star formation, black hole formation, metal cooling,   and background UV/X-ray photoionisation radiation~\cite{Grand:2016mgo}. The  simulations reproduce the observed stellar sizes, masses, rotation curves, star formation rates and metallicities of present day MW-mass galaxies. The Planck-2015~\citep{Planck:2015fie} cosmological parameters are adopted in the simulations: $\Omega_{m}=0.307$, $\Omega_{\rm bar}=0.048$, $H_0=67.77~{\rm km~s^{-1}~Mpc^{-1}}$. We use the standard resolution level (Level 4) of the simulations. At this resolution level, the DM particle mass is $m_{\rm DM} \sim 3\times 10^5~\Msun$, the baryonic particle mass is $m_b=5\times10^4~\Msun$, and the Plummer equivalent gravitational softening length is $\epsilon=370$~pc~\citep{Power:2002sw,Jenkins2013}.

The MW-LMC analogue used in this work is the re-simulated halo 13 in ref.~\cite{Smith-Orlik:2023kyl}, which corresponds to the Auriga 25 halo and its LMC analogue. This system had been re-simulated with finer snapshots close to the LMC's pericenter approach. The MW analogue's virial mass is $1.2 \times 10^{12}~\Msun$ and the LMC analogue's halo mass at infall is $3.2 \times 10^{11}~\Msun$. More details about this system and its selection criteria are given in ref.~\cite{Smith-Orlik:2023kyl}. We consider two snapshots for this halo: isolated MW (\emph{Iso.}) and the present day MW-LMC (\emph{Pres.}). Iso.~takes place at LMC's first apocenter before infall when the MW and LMC analogues have the largest separation, while Pres.~is the snapshot closest to the present day separation of the observed MW and LMC system.  The speed and distance of the LMC analogue with respect to the host MW analogue at the Pres.~snapshot  are 317~km/s and $\sim 50$~kpc, respectively, which closely match  the observed values.

The Sun's position (and velocity) in the simulated halo is chosen such that it matches the observed Sun-LMC geometry, as discussed in detail in ref.~\cite{Smith-Orlik:2023kyl}. In the first step, we identify the stellar disk orientations in the MW analogue that make the same angle with the orbital plane of the LMC analogues as in observations. In the second step, for each allowed disk orientation we find the Sun's position by matching the angles between LMC's orbital angular momentum and the Sun's position and velocity vectors in the simulations to their observed values. In the last step, we determine the best fit Sun's position by selecting the position that results in the closest match of the angles between the Sun's velocity vector and the LMC's position and velocities with their observed values.

We define the \emph{Solar region} as the overlap of the following two regions: a spherical shell between 6 to 10 kpc from the center of the MW analogue (with the Sun at $\sim 8$~kpc from the galactic center) and a cone with an opening angle of $\pi/4$ radians, axis aligned with the best fit Sun's position, and vertex at the galactic center. This region is small enough to be sensitive to the best fit Sun's position, yet large enough to incorporate several thousand DM particles, as detailed in ref.~\cite{Smith-Orlik:2023kyl}. Notice that for the Iso.~snapshot, there is no LMC analogue and we cannot define the best fit Sun's position. Therefore, in this case the Solar region is defined as the spherical shell with radii between 6 to 10~kpc from the center of the MW analogue.

\section{Local dark matter distribution}
\label{sec:fv}

A key ingredient for the computation of the predicted event rate in DM direct detection experiments is the astrophysical contribution (see section~\ref{sec:directdetection}), which encompasses the local DM density and velocity distribution. The local DM density is just a normalization in the event rate, but there is a more complicated dependence of the event rate on the local DM velocity distribution, which enters through an integration. In this section, we present the
DM density and velocity distribution in the Solar region (defined in section~\ref{sec:sims}) extracted from the simulated MW-LMC analogue.

We define the Galactocentric reference frame with the origin on the Galactic center, the $x_g$-axis pointing from the Sun towards the Galactic center, the $y_g$-axis in the direction of the Galactic rotation, and the $z_g$-axis pointing towards the North Galactic Pole. The $x_g$-$y_g$ plane is aligned with the Sun's orbital  plane. We   specify the orientation of the $(x_g, y_g, z_g)$ axes for the best fit Sun's position and velocity, which results in the closest match with the observed Sun-LMC geometry (as discussed in section~\ref{sec:sims}). Next, we  transform the positions and velocities of the simulation particles to this Galactic reference frame.

The local DM density for the Pres.~snapshot is 0.39~{GeV/cm$^3$}, which agrees very well with the local~\cite{Salucci:2010qr, Smith:2011fs, Bovy:2012tw, Garbari:2012ff, Zhang:2012rsb, Bovy:2013raa, Posti:2019, Buch:2018qdr} and global~\cite{McMillan:2011wd, Catena:2009mf, Weber:2009pt, Iocco:2011jz, Nesti:2013uwa, Sofue:2015xpa, Pato:2015dua, deSalas:2019pee} estimates from observations. The percentage of DM particles originating from the LMC in the Solar region\footnote{This percentage is defined as the ratio of the number of DM particles originating from the LMC and the total number of DM particles in the Solar region, multiplied by 100.} is 0.26 for the Pres.~snapshot and zero for the Iso.~snapshot. 

\begin{figure}[t]
    \centering
        \includegraphics[width=9cm]{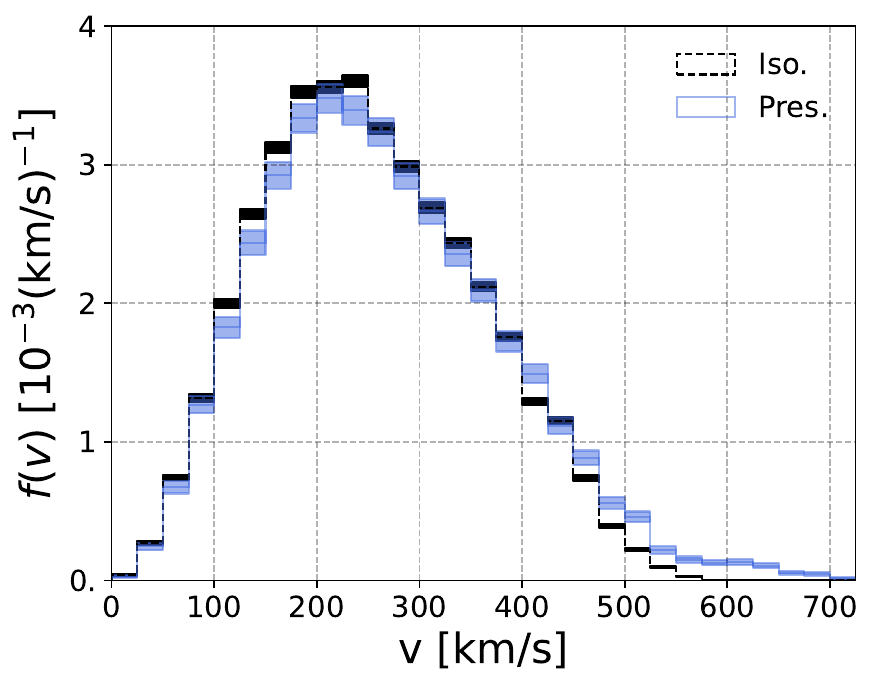} 
    \caption{The local DM speed distribution in the Galactic rest frame in the Solar region for the isolated MW (black) and present day MW-LMC (blue) snapshots. The shaded bands correspond to the 1$\sigma$ Poisson errors in the distributions.}
\label{fig:vel_dist}
\end{figure}

Figure~\ref{fig:vel_dist} shows the local DM speed distribution in the Galactic rest frame for the Iso.~snapshot (black) and the Pres.~snapshot (blue). There are no LMC particles in the Solar region in the Iso.~snapshot. The speed distributions are defined as $f(v) = v^2 \int d\Omega_{\vec v}\Tilde{f}(\vec v)$, where $\Tilde{f}(\vec v)$ is the normalized DM velocity distribution, $d\Omega_{\vec v}$ is an infinitesimal solid angle around the direction $\vec v$, and $\int dv f(v)=\int d^3v \Tilde{f}(\vec v) = 1$. The shaded bands specify the $1\sigma$ Poisson errors in the speed distributions. Since the local circular speed of the MW is commonly set to 220~{\rm km/s} in the SHM, the DM speeds in figure~\ref{fig:vel_dist} are scaled by $(220~{\rm km/s})/v_c$, where $v_c$ is the local circular speed of the MW analogue, calculated from the total mass contained  in a sphere of radius 8~kpc. The impact of the LMC on the high speed tail of the local DM speed distribution is clearly visible at the Pres.~snapshot.

\section{Dark matter direct detection}
\label{sec:directdetection}

In this section we discuss the relevant expressions and details needed to compute the predicted signals in DM direct detection experiments for non-standard DM-nucleus interactions.

\subsection{Halo integrals}
\label{sec:halo}

We consider the case of a DM particle of mass $m_\chi$ scattering with a target nucleus of mass $m_T$ in an underground detector, and depositing the nuclear recoil energy, $E_R$. The differential event rate is given by
\begin{equation}
 \frac{dR}{dE_R}=\frac{\rho_\chi}{m_\chi}\frac{1}{m_T}\,\int_{v>v_{\rm min}} d^3 v\, \frac{d\sigma_T}{dE_R}\,v\, \tilde{f}_{\rm det}({\vec v}, t), 
\label{eq:rate}
\end{equation}
where $\rho_\chi$ is the local DM density, $\sigma_T $ is the DM-nucleus scattering cross section, $\vec v$ is the relative velocity between the DM and target nucleus, with $v=|{\vec v}|$, $\tilde{f}_{\rm det}({\vec v}, t)$ is the local DM velocity distribution in the detector reference frame, and $v_{\rm min}$ is the minimum speed needed for the DM particle to impart a recoil energy $E_R$ in the detector.

For the case of elastic scattering, which we consider in section~\ref{sec:NREFT}, we have
\begin{equation}
 v_{\rm min}=\sqrt{\frac{m_T E_R}{2\mu_{\chi T}^2}}\,,   
\end{equation}
where $\mu_{\chi T}$ is the DM-nucleus reduced mass. 

For the case of inelastic DM scattering~\cite{Tucker-Smith:2001myb}, the DM particle $\chi$ scatters to an excited state $\chi^\ast$, with a mass difference $\delta=m_\chi^\ast-m_\chi$. We consider endothermic scattering where $\delta>0$. We have
\begin{equation}
 v_{\rm min}=\sqrt{\frac{1}{2m_T E_R}}\left(\frac{m_T E_R}{\mu_{\chi T}}+\delta\right).
\label{eq:vmin_inel}
\end{equation}
The dependence of $v_{\rm{min}}$ on the mass splitting, $\delta$, increases the minimum DM speed required for the DM particle to produce a recoil signal. We present the results for this case in section~\ref{sec:results}.

For the standard spin-independent and spin-dependent DM-nucleus interactions, the differential cross section is proportional to $v^{-2}$ and the  event rate becomes proportional to the {\it halo integral},
\begin{equation}
    \eta(v_{\rm{min}},t) \equiv \int_{v > v_{\rm{min}}} d^3v \, \frac{\tilde{f}_{\rm{det}}({\vec v},t)}{v},
\label{eq:etavmin}
\end{equation}
which along with the local DM density, $\rho_\chi$, encorporates the astrophysical dependence of the event rate.

For the case of general non-standard interactions, such as the set of effective operators discussed in section~\ref{sec:NREFT}, the differential cross section may have a different dependence on the relative velocity between the DM and the nucleus, and also vary with the exchanged momentum. In particular, for a very general set of non-relativistic effective operators, the differential cross section can be expressed as a linear combination of a velocity-dependent term and a velocity-independent term~\cite{Kahlhoefer:2017ddj, Cirelli:2013ufw},
\begin{equation}
\frac{d\sigma_T}{dE_R} = \frac{d\sigma_1}{dE_R}\frac{1}{v^2} + \frac{d\sigma_2}{dE_R}.
\label{eq:gencrosssection}
\end{equation}
The first term results in the halo integral, $\eta(v_{\rm min}, t)$, given in eq.~\eqref{eq:etavmin}, while the second term results in a different velocity integral defined as
\begin{equation}
    h(v_{\rm{min}},t) \equiv \int_{v>v_{\rm{min}}} d^3 v\, v\, \tilde{f}_{\rm{det}}({\vec v},t).
\label{eq:hvmin}
\end{equation}
The differential rate in eq.~\eqref{eq:rate} can then be written as
\begin{equation}
\frac{dR}{dE_R} = \frac{\rho}{m_\chi}\frac{1}{m_T} \left[ \frac{d\sigma_1}{dE_R}\,\eta (v_{\rm{min}}, t) + \frac{d\sigma_2}{dE_R}\,h(v_{\rm{min}}, t)\right].
\label{eq:genrate}
\end{equation} 
When the last term in eq.~\eqref{eq:genrate} is absent, we recover the standard interaction and all the astrophysical dependence of the event rate is contained solely in the halo integral, $\eta(v_{\rm{min}}, t)$. In section~\ref{sec:NREFT} we discuss the effective operators which can lead to non-standard interactions.

The LMC's influence on the local DM velocity distribution will lead to a direct impact on both $\eta(v_{\rm min}, t)$ and $h(v_{\rm min}, t)$ and thus, on the predicted signals in direct detection experiments. In order to compute these quantities, we boost the local DM velocity distribution of the simulated halo from the galactic reference frame to the detector frame, by $\tilde{f}_{\rm{det}}({\vec v},t)=\tilde{f}_{\rm{gal}}({\vec v} + {\vec v}_s + {\vec v}_e(t))$. Here ${\vec v}_s={\vec v}_c + {\vec v}_{\rm{pec}}$ is the velocity of the Sun in the Galactic rest frame, where $\vec v_c$ is the Sun's circular velocity (which we take as $v_c=|\vec v_c|=220$~km/s) and ${\vec v}_{\rm{pec}}=(11.10,12.24,7.25)$~km/s~\cite{2010MNRAS.403.1829S} is the Sun's peculiar velocity in Galactic coordinates. ${\vec v}_e(t)$ is the velocity of the Earth with respect to the Sun. We neglect the small eccentricity of the Earth's orbit, for simplicity.

\begin{figure*}[t]
    \centering
    \begin{subfigure}
        \centering
        \includegraphics[width=7.5cm]{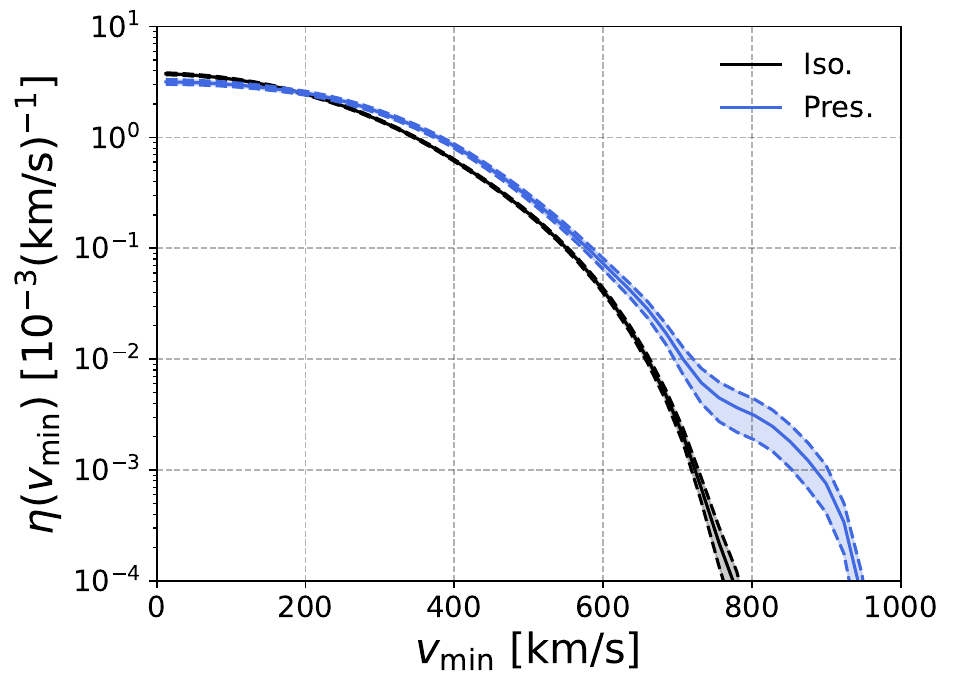}
    \end{subfigure}%
    ~ 
    \begin{subfigure}
        \centering
        \includegraphics[width=7.5cm]{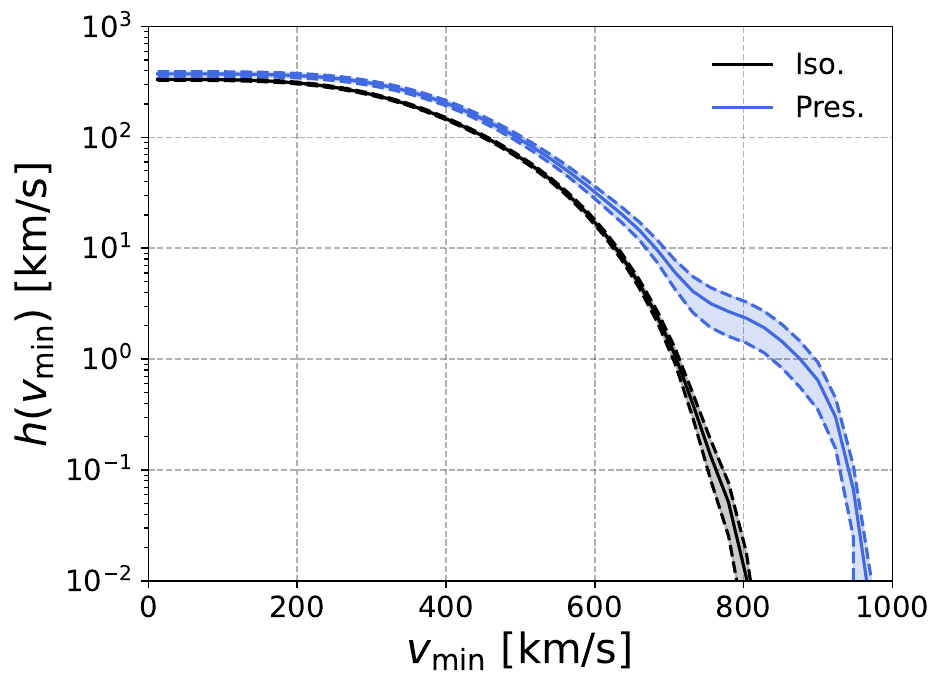}
    \end{subfigure}
    \caption{Time averaged halo integrals $\eta(v_{\rm min})$ (left panel) and $h(v_{\rm min})$ (right panel). The black and blue solid lines correspond to the halo integrals obtained from the mean local DM velocity distribution for the isolated MW and present day MW-LMC snapshots, respectively. The
shaded bands are obtained from the local DM velocity distribution at $1\sigma$ from the mean in each case.}
\label{fig:halo_integrals}
\end{figure*}

We show the time averaged halo integrals $\eta(v_{\rm min})$ (eq.~\eqref{eq:etavmin}) and $h(v_{\rm min})$ (eq.~\eqref{eq:hvmin}) in the left and right panels of  figure~\ref{fig:halo_integrals}, respectively. The black and blue solid curves are the halo integrals computed from the mean value of the local DM velocity distributions shown in figure~\ref{fig:vel_dist} for the Iso.~and Pres.~snapshots, respectively. In both cases the shaded bands show the $1\sigma$ uncertainty in the halo integrals and are obtained from the DM velocity distribution at $1\sigma$ from the mean. We can see that the LMC significantly impacts the high speed tails of both halo integrals in the Pres.~snapshot, leading to shifts of more than 150~km/s in $\eta({v_{\rm min}})$ and more than 160~km/s in $h(v_{\rm min})$.

\subsection{Non-relativistic effective operators}
\label{sec:NREFT}

The NREFT approach describes all
possible contact DM-nucleon interactions arising from a full theory of DM~\cite{Fan:2010gt, Fitzpatrick:2012ix, Fitzpatrick:2012ib, Anand:2013yka, Dent:2015zpa}. This formalism extends  the standard spin-independent and spin-dependent interactions, and includes velocity-dependent and momentum-dependent operators. In this framework, the DM-nucleus interaction is parametrized in terms of a set of independent operators, $\mathcal{O}_i$, listed in the first column of table~\ref{tab:operators}. In this notation, $\vec q$ is the momentum transferred from the DM to the nucleon, $\vec v_\perp$ is the component
of the relative velocity between the DM and the
nucleon that is perpendicular to $\vec q$, $\vec S_\chi$ is the DM spin operator, $\vec S_N$ is the nucleon spin operator, and $m_N$ is the nucleon mass.

\renewcommand{\arraystretch}{2}
\begin{table}
\centering
\begin{tabular}{|c|c|}
\hline
{Operator} & {Scaling factor}   \,  \\ 
 \hline
       $\mathcal{O}_1=1_\chi 1_N$ & 1  \\
       $\mathcal{O}_3=i \vec S_N \cdot \left(\frac{\vec q}{m_N} \times \vec v_{\perp}\right)$ &  $q^2 v_\perp^2$, $q^4$ \\
       $\mathcal{O}_4=\vec S_\chi \cdot \vec S_N$ & 1  \\
       $\mathcal{O}_5=i \vec S_\chi \cdot \left(\frac{\vec q}{m_N} \times \vec v_{\perp}\right)$ & $q^2\,  v_\perp^2$, $q^4$  \\
       $\mathcal{O}_6=\left(\vec S_\chi \cdot \frac{\vec q}{m_N} \right) \left(\vec S_N \cdot \frac{\vec q}{m_N} \right)$  & $q^4$  \\
       $\mathcal{O}_7=\vec S_N \cdot \vec v_\perp$ & $v_\perp^2$   \\
       $\mathcal{O}_8=\vec S_\chi \cdot \vec v_\perp$ & $v_\perp^2$, $q^2$ \\
       $\mathcal{O}_9=i \vec S_\chi \cdot \left(\vec S_N \times \frac{\vec q}{m_N}\right)$ & $q^2$  \\
       $\mathcal{O}_{10}=i \vec S_N \cdot \frac{\vec q}{m_N}$ & $q^2$  \\
       $\mathcal{O}_{11}=i \vec S_\chi \cdot \frac{\vec q}{m_N}$ & $q^2$ \\
       $\mathcal{O}_{12}=\vec S_\chi \cdot \left(\vec S_N \times \vec v_\perp \right)$ & $v_\perp^2$, $q^2$ \\
       $\mathcal{O}_{13}=i\left(\vec S_\chi \cdot \vec v_\perp \right) \left(\vec S_N \cdot \frac{\vec q}{m_N} \right)$ & $q^2 v_\perp^2$, $q^4$  \\
       $\mathcal{O}_{14}=i\left(\vec S_\chi \cdot \frac{\vec q}{m_N} \right) \left(\vec S_N \cdot \vec v_\perp \right)$ & $q^2 v_\perp^2$  \\
       $\mathcal{O}_{15}=-\left(\vec S_\chi \cdot \frac{\vec q}{m_N} \right) \left(\left(\vec S_N \times \vec v_\perp \right)\cdot \frac{\vec q}{m_N}\right)$ & $q^4 v_\perp^2$, $q^6$  \\
\hline
\end{tabular}
\caption{\label{tab:operators} The independent operators that parametrize the DM-nucleus interaction in the NREFT approach (first column), and the scaling of the differential DM-nucleus cross section in powers of $q^2$ and $v_\perp^2$ that appear in the DM response functions, $R_k$, relative to the standard spin-independent $\mathcal{O}_1$ operator (second column).}
\end{table}

This formalism considers one-body DM-nucleon interactions mediated by a heavy spin-1 or spin-0 particle. We consider a spin-$1/2$ DM particle, and  assume isospin independent interactions, i.e.~equal couplings for protons and neutrons. We further neglect interference between different operators. The DM-nucleus differential cross section for an elastic DM interaction can then be factorized in terms of the DM response functions and the nuclear response functions~\cite{Catena:2015uha, Anand:2013yka}. For isoscalar interactions, the differential cross section for the DM scattering with a target nucleus $T$ is given by
\begin{align}
\frac{d\sigma_T}{d E_R}=&\frac{2 m_T}{v^2 (2J+1)}\left[\sum_{k=M, \Sigma',\Sigma''} R_k(v_\perp^2, q^2)\, W_k (q^2)+\frac{q^2}{m_N^2}\sum_{k=\Phi'', \tilde{\Phi}', \Delta} R_k(v_\perp^2, q^2)\, W_k (q^2) \right],
\label{eq:crosssec}
\end{align}
where $J$ is the nuclear spin. The $k$ index in the two sums appearing in eq.~\eqref{eq:crosssec} extends over the six DM response functions, $R_k$, and the six nuclear response functions, $W_k$. Notice that since we are only considering isoscalar interactions we have dropped the superscripts $\tau=\tau'=0$ from the usual notation of the DM and nuclear response functions and have set $R_k^{00}\equiv R_k$ and $W_k^{00} \equiv W_k$. Notice also that there are two mixed DM response functions, $R_{\Phi'' M}$ and $R_{\Delta \Sigma'}$, which only appear when the couplings to at least two operators are nonzero. Since we are neglecting the combination of different operators and consider the couplings to be nonzero one at a time, these two mixed response functions are not included in eq.~\eqref{eq:crosssec} and not considered in this work.

The DM response functions  depend on the DM-nucleus relative velocity, the momentum transfer, and the DM-nucleon interaction strength. The nuclear response functions for a given nucleus are functions of only $q^2$. These response functions are provided in the appendices of refs.~\cite{Catena:2015uha,Catena:2015vpa}. The differential cross section can be classified according to the powers of $q^2$ and $v_\perp^2$ that appear in the DM response functions. In the second column of table~\ref{tab:operators}, we list the dependence of the differential cross section on powers of $q^2$ and $v_\perp^2$ appearing in the DM response functions, $R_k$, relative to the standard spin-independent $\mathcal{O}_1$ operator.

The experimental signatures of different NREFT operators sensitively depend on the properties of the target nucleus in the detector, as well as the DM halo. Multiple experiments have already published constraints on the WIMP-nucleon effect couplings~\cite{DarkSide-50:2020swd,LUX:2021ksq,SuperCDMS:2022crd,SuperCDMS:2015lcz,DEAP:2020iwi, LZ:2023lvz}.

We would like to investigate how the LMC impacts the  exclusion limits set by different near-future direct detection experiments on the couplings to the  operators listed in table~\ref{tab:operators}. To directly compare the results for different operators and experiments, we follow the notation of ref.~\cite{DEAP:2020iwi}, and define the effective DM-proton cross section
\begin{equation}
\sigma_{\chi p} \equiv \frac{\left( c_{i}^{p}\, \mu_p \right)^{2}}{\pi},
\label{eq:sigma_ci}
\end{equation}
where $\mu_p$ is the DM-proton reduced mass and $c_i^p$ is the effective DM-proton coupling. For isoscalar interactions, 
\begin{equation}
c_{i}^{p} = c_{i}^{n}= c_i^0/2,
\end{equation}
where $c_i^0$ is the coupling to each operator $\mathcal{O}_i$ and has dimension of $1/{\rm (mass)}^2$.

Notice that eq.~\eqref{eq:sigma_ci} leads to the standard spin-independent DM-nucleon cross section for  the $\mathcal{O}_1$ operator, but may not  directly correspond to a physical cross section for the other operators.

\section{Experiments}
\label{sec:experiments}

In this section, we provide the experimental parameters used to calculate the expected exclusion limits set by six near-future experiments: DarkSide-20k, SBC, DARWIN$\slash$XLZD, SuperCDMS, NEWS-G and DarkSPHERE. These experiments use different detector technologies and target nuclei as well as different nuclear energy thresholds. DarkSide-20k~\cite{DarkSide-20k:2017zyg} and DARWIN$\slash$XLZD~\cite{DARWIN:2016hyl, Aalbers:2022dzr, Baudis:2024jnk} detectors consist of double-phase time projection chambers filled with argon and xenon, respectively. The SBC experiment \cite{Alfonso-Pita:2023frp} uses a bubble chamber detector technology filled with superheated liquid argon. SuperCDMS~\cite{SuperCDMS:2016wui} detectors use germanium and silicon cryogenic crystals cooled at millikelvin temperatures. Finally, NEWS-G~\cite{NEWS-G:2022kon} and DarkSPHERE are spherical proportional counters filled with noble-gas mixtures such as neon and helium, respectively. 

We use the publicly available code \texttt{DDCalc} \cite{GAMBITDarkMatterWorkgroup:2017fax} to compile the results by considering the exposure, number of background and expected events, and nuclear recoil efficiencies from the following publications for consistency:  Ref.~\cite{DarkSide-20k:2017zyg} is used for DarkSide-20k, refs. \cite{Alfonso-Pita:2023frp, Bressler:2022} for SBC, ref.~\cite{Schumann:2015cpa} for DARWIN$\slash$XLZD, ref.~\cite{SuperCDMS:2016wui} for SuperCDMS with germanium crystal, ref.~\cite{Durnford:2021mzg} for NEWS-G with neon and ref.~\cite{NEWS-G:2023qwh}  for DarkSPHERE with helium. In the following, we provide the specifications considered for each experiment.

\begin{itemize}
  \item \textbf{DarkSide-20k.} We use a total exposure of $3.65\times 10^{7}$ $\rm{kg}\cdot\rm{day}$ with a total expected number of events of 0 and 1 $\nu$-induced background event. The nuclear recoil efficiency curve is taken from figure~92 from ref.~\cite{DarkSide-20k:2017zyg} within an energy range of [$30 - 200$] keVnr.
  
  \item \textbf{SBC.} We use a total exposure of $3.65\times 10^{3}$ $\rm{kg}\cdot\rm{day}$ with a total expected number of events of 0 and 2.46 $\nu$-induced background event \cite{Bressler:2022}. The nuclear recoil efficiency curve is taken from figure~8 from ref.~\cite{Li:2024xzv} over an energy range of [$0.1 - 10$] keVnr.  
  
  \item \textbf{DARWIN$\slash$XLZD.} Due to the complexity of the analysis method to calculate the expected cross section limits in recent publications, such as ref.~\cite{Aalbers:2022dzr} often used also for DARWIN$\slash$XLZD, we decided to use the already built-in module in \texttt{DDCalc}, which follows the specifications from ref.~\cite{Schumann:2015cpa}. The total exposure considered is $7.3\times 10^{7}$ $\rm{kg}\cdot \rm{day}$, and the total expected number of events is 1 with 2.37 background events. The nuclear recoil efficiency is constant with an acceptance of $30\%$ over an energy range of [$5 - 21$] keVnr.

  \item \textbf{SuperCDMS.} We consider the high-voltage (HV) germanium detector with a total exposure of 1.6 $\times 10^{4}$ $\rm{kg}\cdot\rm{day}$. We take a constant efficiency of  85\% acceptance and add an extra cut of $50\%$ for the radial fiducial volume, as detailed in refs.~\cite{SuperCDMS:2016wui,SuperCDMS:2015eex}, within an energy range of [$0.04 - 0.3$] keVnr. For the background model, we use the differential rate from \cite{SuperCDMS:2016wui} to be a constant value of 10 $\rm{keV}^{-1} \rm{kg}^{-1} \rm{yr}^{-1}$, which is a good approximation in the energy range of consideration \cite{Smith-Orlik:2023kyl,Kahlhoefer:2017ddj}. For the analysis, we assume 51 total expected events with no background event. 

  \item \textbf{NEWS-G.} We consider only neon as the active target with a total exposure of 18~$\rm{kg}\cdot\rm{day}$. We use a flat background of 1.65 $\rm{keV}^{-1} \rm{kg}^{-1} \rm{yr}^{-1}$ \cite{Durnford:2021mzg}, which gives a total expected events number of 30 with no background event. The nuclear recoil efficiency is assumed to be constant with $100\%$ acceptance over an energy range of $[0.03 - 1]$~keVnr.

  \item \textbf{DarkSPHERE.} We consider only helium ($^{4}{\rm He}$ isotope) as the active target with a total exposure of $7.4\times 10^{3}$ $\rm{kg}\cdot\rm{day}$. We use a flat background of 0.01 $\rm{keV}^{-1} \rm{kg}^{-1} \rm{yr}^{-1}$, which gives a total expected events number of 70 with no background event \cite{NEWS-G:2023qwh}. As NEWS-G, the nuclear recoil efficiency is assumed to be constant with $100\%$ acceptance within an energy range of [$0.03 - 1$] keVnr. 
  
\end{itemize}

For the NEWS-G and DarkSPHERE experiments, the efficiencies are usually calculated in terms of the electron equivalent energy scale (keVee) instead of the nuclear recoil energy scale (keVnr). The conversion between these two scales is not trivial and requires several analysis steps to obtain the energy response in terms of keVnr. Using \texttt{DDCalc}, the calculation of the cross section requires the nuclear recoil efficiency. By assuming a nuclear recoil efficiency to be constant with $100\%$ acceptance within an energy range of [$0.03 - 1$] keVnr, we found comparable results with the published projection limits at low WIMP masses for NEWS-G ($\leq$ 500 MeV) \cite{Durnford:2021mzg} and DarkSPHERE ($\leq$ 400 MeV) \cite{NEWS-G:2023qwh}. For higher WIMP masses, there is a constant deviation but staying within the same order of magnitude.  Since the impact of the LMC mostly affects the low WIMP mass range, we consider that the nuclear recoil efficiency assumed in our analysis is a good approximation for reproducible results.

Table \ref{table:experiments} summarizes some of the relevant quantities we  outlined above for the different experiments.

\begin{table}[t]
\centering
\begin{tabular}{|c c c c |}
 \hline
 Experiment & Target Nucleus & Exposure [kg$\cdot$ day] & Energy range [keVnr] \\ [0.5ex] 
 \hline%\hline 
 DarkSide-20k \cite{DarkSide-20k:2017zyg} & Ar & $3.65\times 10^{7}$ & [$30 - 200$] \\
  \hline
 SBC \cite{Alfonso-Pita:2023frp} & Ar & $3.65\times 10^{3}$ & [$0.1 - 10$] \\
 \hline
 DARWIN$\slash$XLZD \cite{Schumann:2015cpa} & Xe & $7.3\times 10^{7}$ & [$5 - 21$] \\
 \hline
  SuperCDMS \cite{SuperCDMS:2016wui} & Ge & $1.6 \times 10^{4}$ & [$0.04 - 0.3$]  \\ 
 \hline
 NEWS-G \cite{Durnford:2021mzg} & Ne & 18 & [$0.03 - 1$]  \\
 \hline
 DarkSPHERE \cite{NEWS-G:2023qwh} & $^4{\rm He}$ & $7.4\times 10^{3}$ & [$0.03 - 1$]  \\ 
 \hline
\end{tabular}
\caption{\label{table:experiments}Experimental specifications we  implement in our analysis (see section \ref{sec:results}). The columns specify i) the experiment under consideration, ii) the target nucleus we consider, iii) the total exposure in $\rm{kg}\cdot\rm{day}$, and iv) the nuclear recoil energy range considered for each experiment in keVnr.}
\end{table}

\section{Results}
\label{sec:results}

In this section we present the constraints on the effective DM-nucleon cross section (eq.~\eqref{eq:sigma_ci}) set by the six near-future experiments described in section~\ref{sec:experiments}, for each of the NREFT operators given in table~\ref{tab:operators}, as well as for the case of spin-independent inelastic DM scattering. In each case, we present and compare the results using the local DM velocity distribution of the isolated halo (with no LMC particles) and the present day halo (with DM particles from both the MW and LMC).

\begin{figure}[t]
            \includegraphics[width=.5\linewidth]{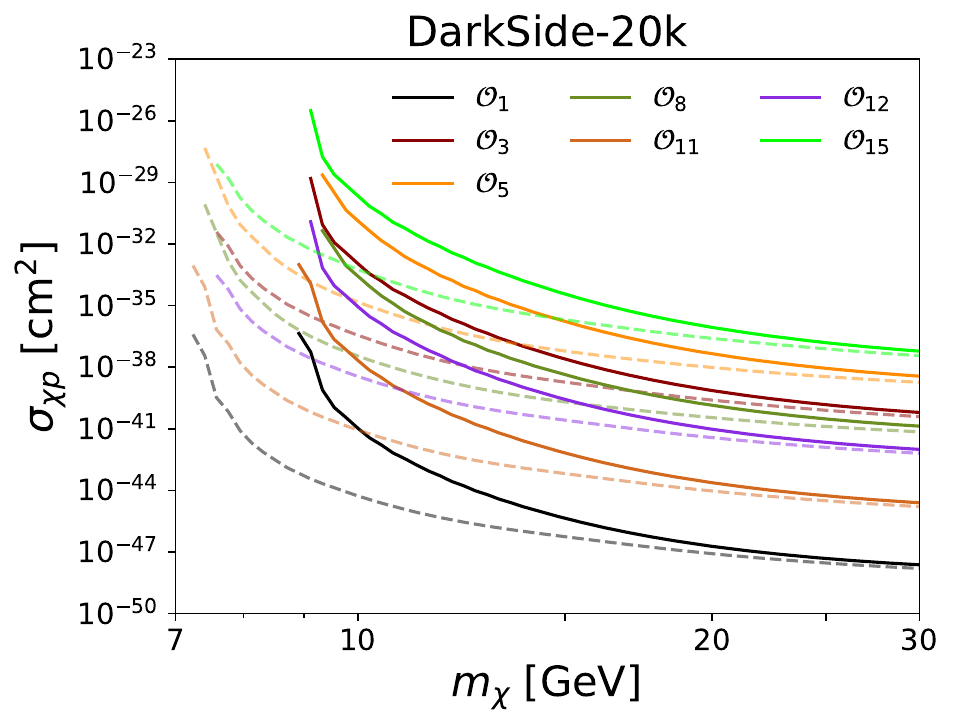}%
        \hfill
            \includegraphics[width=.5\linewidth]{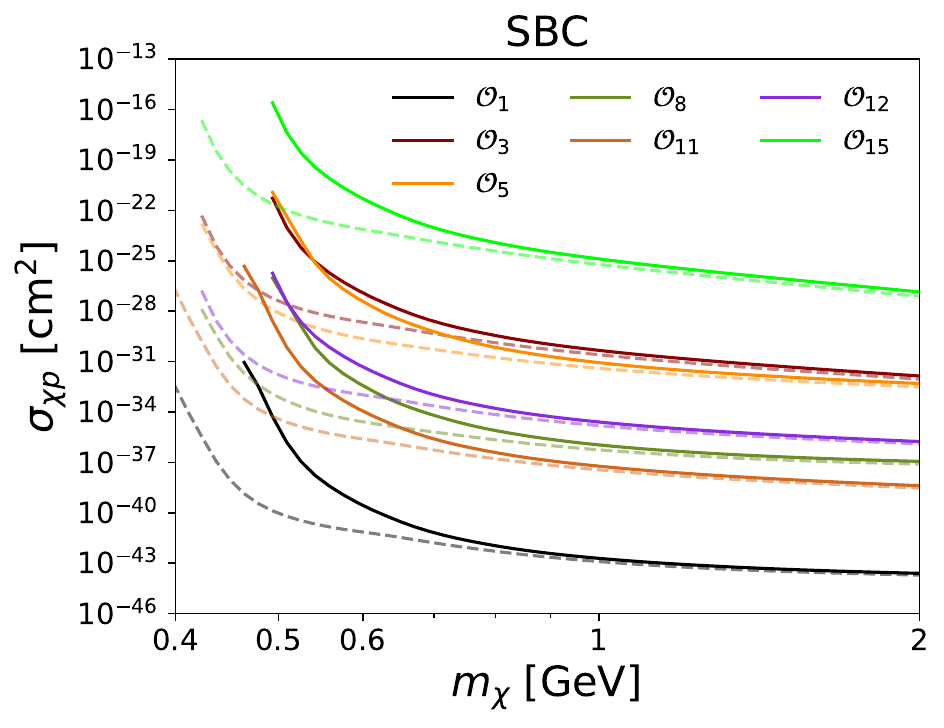}
        \\
            \includegraphics[width=.5\linewidth]{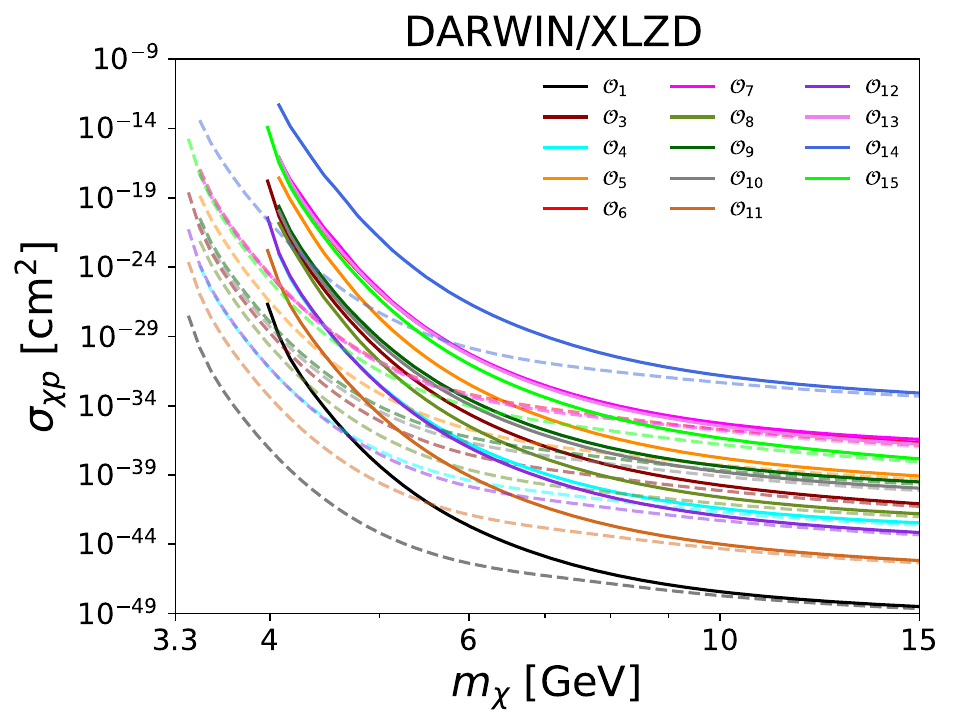}%
        \hfill
            \includegraphics[width=.5\linewidth]{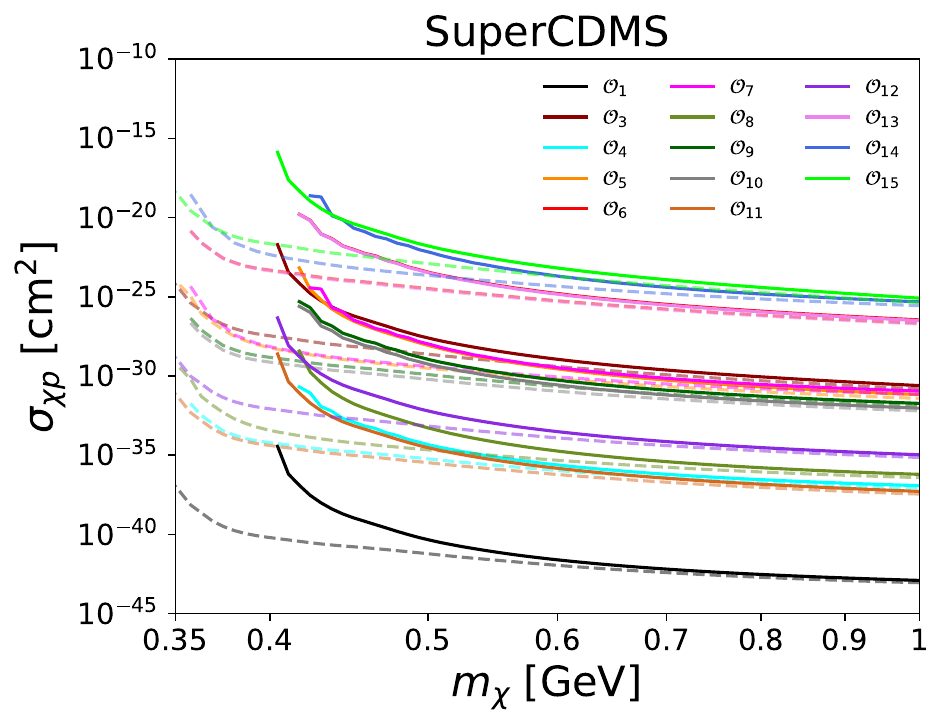}
        \\
             \includegraphics[width=.5\linewidth]{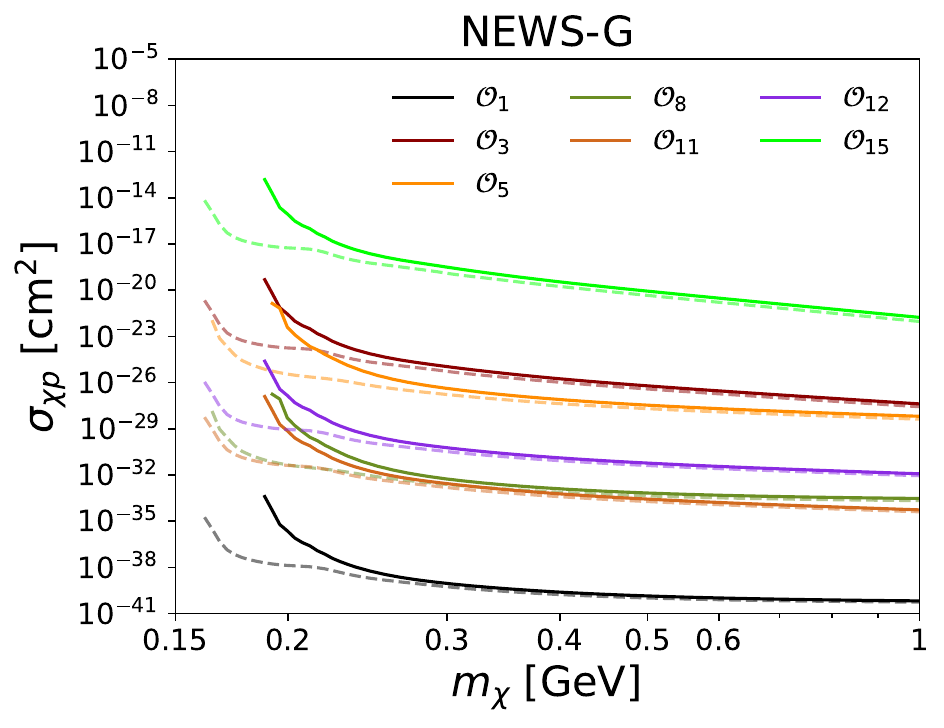}%
        \hfill
            \includegraphics[width=.5\linewidth]{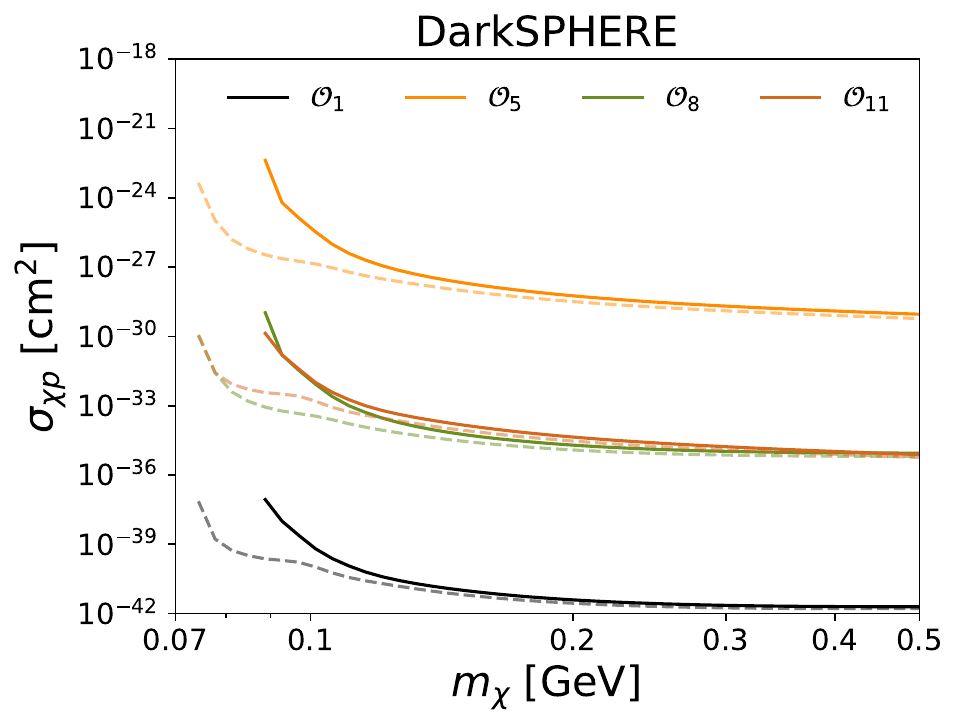}
        
        \caption{Exclusion limits at the 90\% CL in the DM mass and the effective DM-nucleon cross section plane for DarkSide-20k (top left panel),  SBC (top right panel), DARWIN$\slash$XLZD (middle left panel), SuperCDMS (middle right panel), NEWS-G (bottom left panel), and DarkSPHERE (bottom right panel) for the different NREFT operators, $\mathcal{O}_i$, discussed in section~\ref{sec:NREFT}. Solid colored lines are computed using the local DM velocity distribution of the isolated MW analogue, while dashed colored lines are computed for the present day MW-LMC analogue. The local DM density is set to $\rho_\chi = 0.3$~GeV/cm$^3$. 
        }
        \label{fig:all_op}
\end{figure}

Figure \ref{fig:all_op} shows the expected exclusion limits at the 90\% CL in the plane of DM mass, $m_\chi$, and the effective DM-nucleon cross section, $\sigma_{\chi p}$, set by DarkSide-20k (top left panel), SBC (top right panel), DARWIN$\slash$XLZD (middle left panel), SuperCDMS (middle right panel), NEWS-G (bottom left panel), and DarkSPHERE (bottom right panel) for the NREFT operators. Solid lines show the limits corresponding to the local DM velocity distribution of the isolated halo (Iso.), while dashed lines show the limits for the local DM velocity distribution of the present day halo with MW and LMC DM particles (Pres.). The limits have been computed from the mean of the halo integrals (see figure~\ref{fig:halo_integrals}). We have set the local DM density to $\rho_\chi = 0.3$~GeV/cm$^3$ in all panels, to differentiate the impact of the local DM velocity distribution.

Notice that not all operators are present in figure \ref{fig:all_op} for the case of DarkSide-20k, SBC, NEWS-G, and DarkSPHERE since Ar, Ne, and $^4{\rm He}$ have zero spin. This leads to zero event rates for some of the operators that depend on $\vec S_N$, and for cases in which the differential cross section depends on the nuclear response functions $W_{\Sigma'}$ and $W_{\Sigma''}$, which measure the nucleon spin content of the nucleus. However, $\mathcal{O}_3$, $\mathcal{O}_{12}$ and $\mathcal{O}_{15}$, which contain $\vec{S}_N$, can still produce a nuclear recoil in Ar and Ne, since for these operators the nuclear response function $W_{\phi''}$ is sensitive to the nucleon spin-orbit coupling. For $^4{\rm He}$, the $W_{\phi''}$ nuclear response function does not appear, and the only operators present are therefore $\mathcal{O}_1$, $\mathcal{O}_{5}$, $\mathcal{O}_{8}$ and $\mathcal{O}_{11}$.

As seen in figure~\ref{fig:all_op}, the exclusion limits are shifted towards smaller DM masses and smaller cross sections due to the impact of the LMC. This trend exists for all the operators and experiments we consider, and is more pronounced at small DM masses. This is expected, since the variations in the tails of the halo integrals at high $v_{\rm min}$ (shown in figure~\ref{fig:halo_integrals}) lead to the differences in the exclusion limits at low DM masses. This also agrees with the results of refs.~\cite{Smith-Orlik:2023kyl, Besla:2019xbx} for the standard spin-independent interaction. The LMC, therefore, increases the sensitivity of direct detection experiments at low DM masses.

\begin{figure*}[t]
    \centering
    \begin{subfigure}
        \centering
        \includegraphics[width=7.32cm]{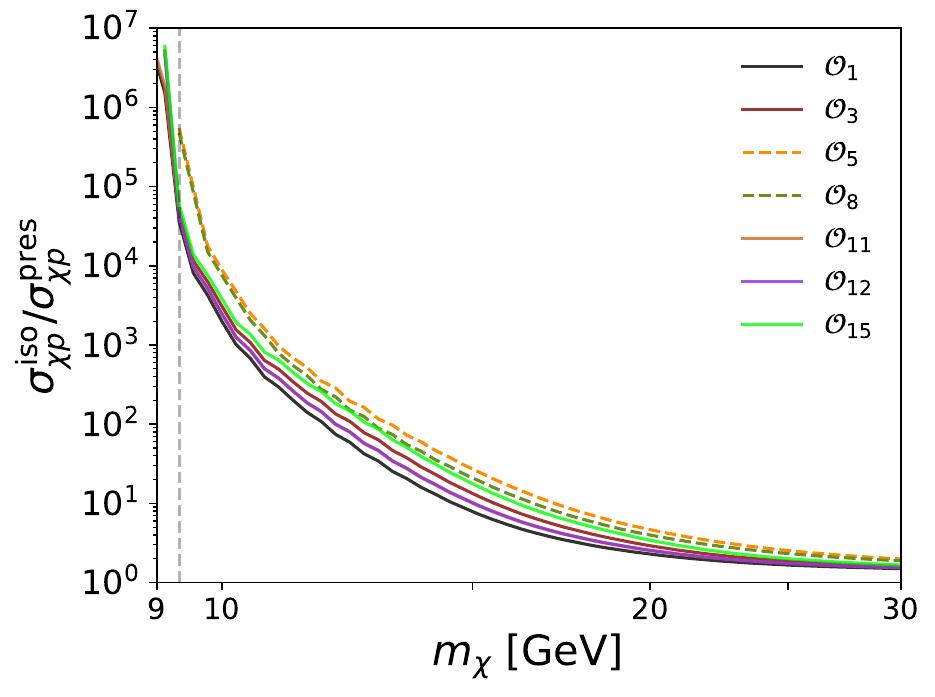}
    \end{subfigure}%
    ~ 
    \begin{subfigure}
        \centering
        \includegraphics[width=7.5cm]{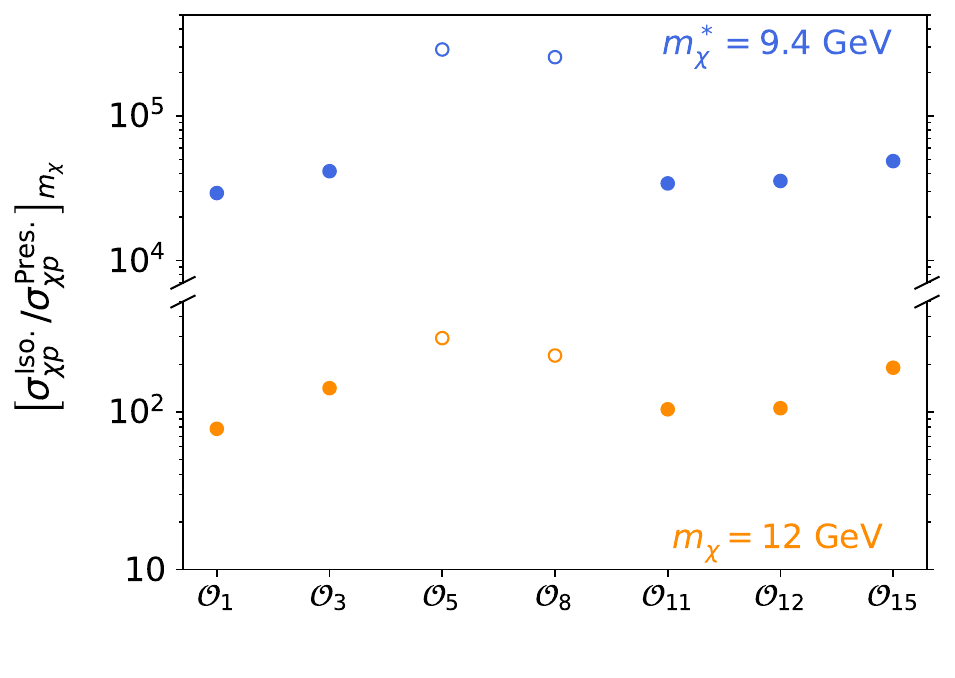}
    \end{subfigure}
    \caption{Left panel: The ratio of the exclusion limit at the 90\% CL on the effective DM-nucleon cross section obtained using the local DM distribution of the isolated MW, $\sigma_{\chi p}^{\rm{iso}}$, and that of the present day MW-LMC, $\sigma_{\chi p}^{\rm{pres}}$, for DarkSide-20k and for different NREFT operators. Each ratio curve starts at the lowest DM mass where the ratio is finite. Right panel: The ratio $\sigma_{\chi p}^{\rm{iso}}/\sigma_{\chi p}^{\rm{pres}}$ evaluated at two fixed DM masses. The blue circles show the ratios evaluated at the minimum DM mass, $m_{\chi}^\ast$, for which $\sigma_{\chi p}^{\rm{iso}}$ is finite for all operators. The value of $m_{\chi}^\ast$ is specified with a gray dashed vertical line in the left panel and given as a legend in the top corner of the right panel. The orange circles show the ratios evaluated at a slightly larger DM mass, whose value is specified in the bottom corner of the right panel. The dashed (solid) colored lines in the left panel and the open (filled) circles in the right panel correspond to operators for which the DM-nucleus cross section has a velocity-dependent (velocity-independent) scaling compared to the $\mathcal{O}_1$ operator. 
    }
\label{fig:darkside}
\end{figure*}

\begin{figure*}[h!]
    \centering
    \begin{subfigure}
        \centering
        \includegraphics[width=7.32cm]{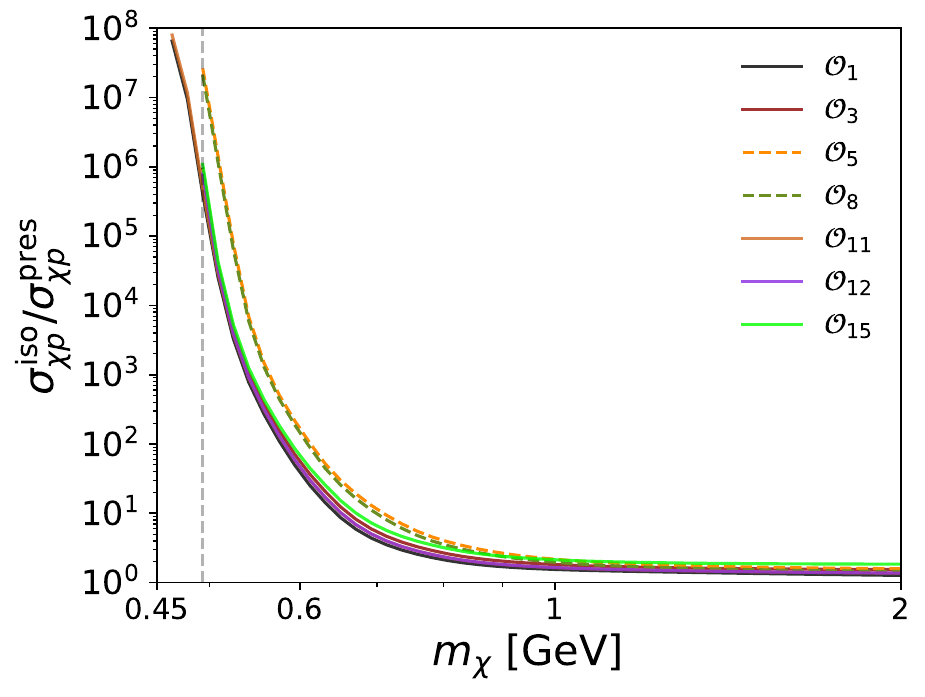}
    \end{subfigure}%
    ~ 
    \begin{subfigure}
        \centering
        \includegraphics[width=7.6cm]{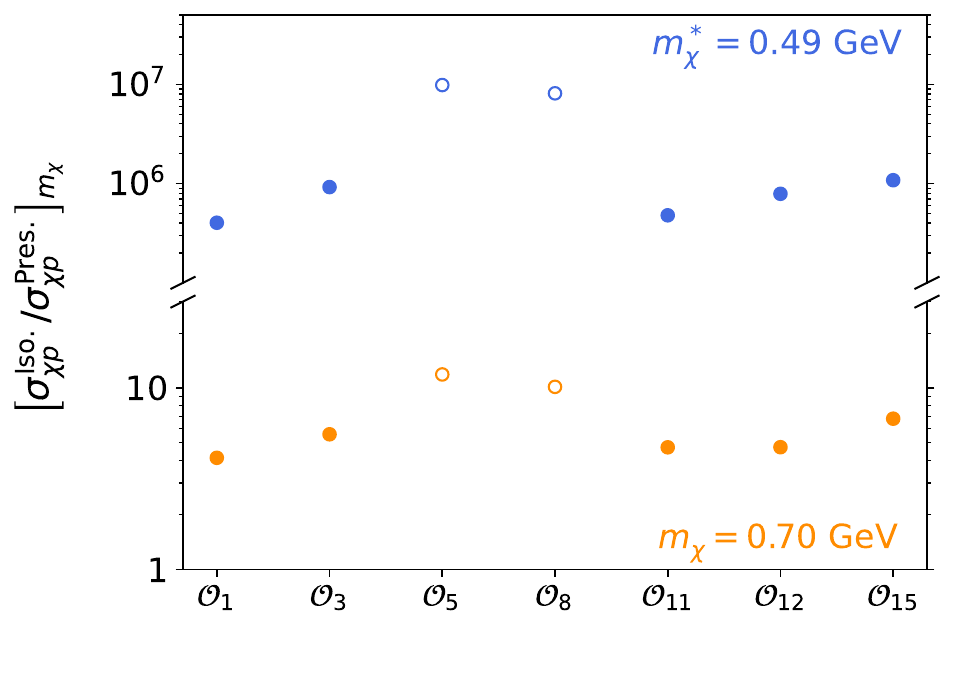}
    \end{subfigure}
    \caption{Same as figure~\ref{fig:darkside}, but for the SBC experiment.}
\label{fig:sbc}
\end{figure*}

\begin{figure*}[h!]
    \centering
    \begin{subfigure}
        \centering
        \includegraphics[width=7.55cm]{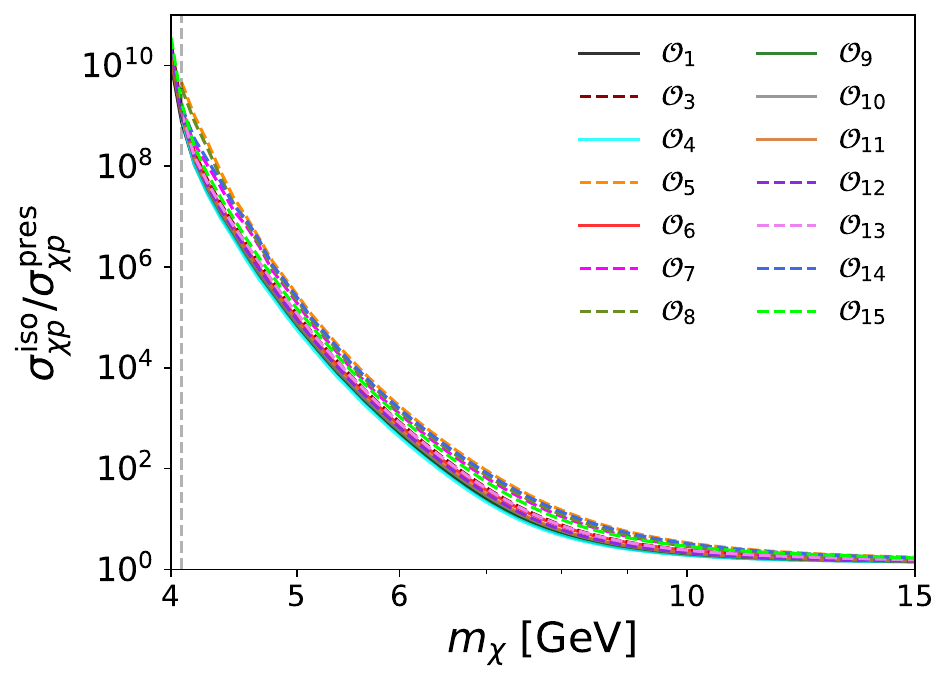}
    \end{subfigure}%
    ~ 
    \begin{subfigure}
        \centering
        \includegraphics[width=7.55cm]{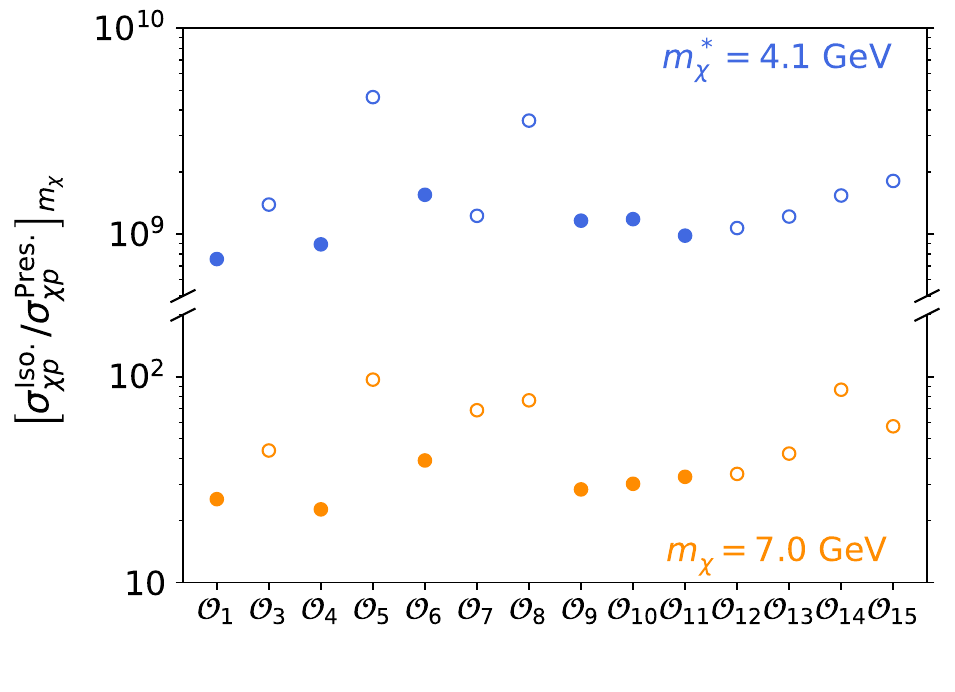}
    \end{subfigure}
    \caption{Same as figure~\ref{fig:darkside}, but for the DARWIN$\slash$XLZD experiment.}
\label{fig:darwin}
\end{figure*}

\begin{figure*}[h!]
    \centering
    \begin{subfigure}
        \centering
        \includegraphics[width=7.32cm]{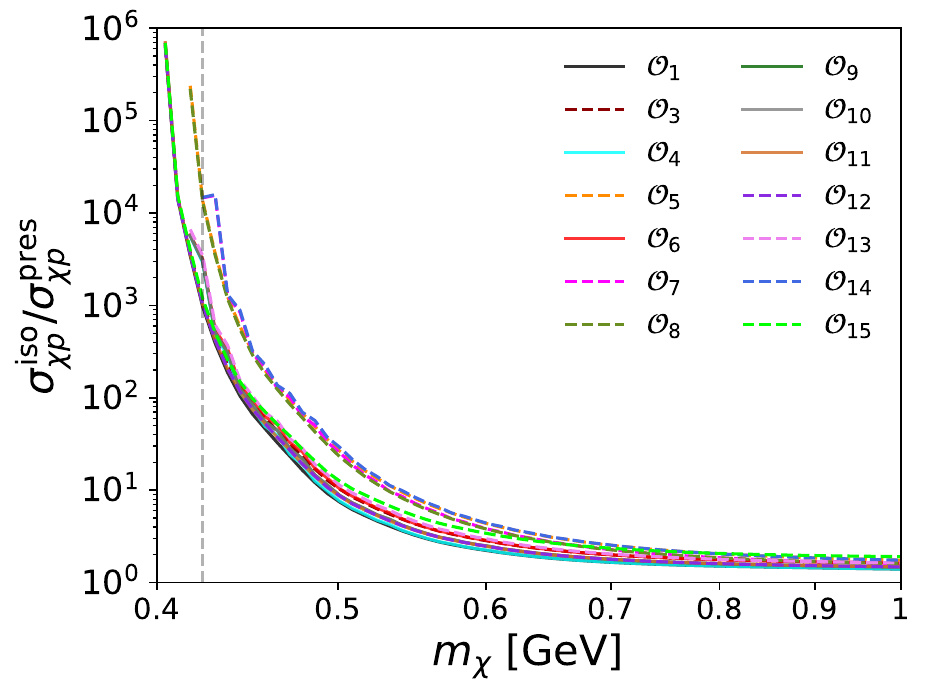}
    \end{subfigure}%
    ~ 
    \begin{subfigure}
        \centering
        \includegraphics[width=7.6cm]{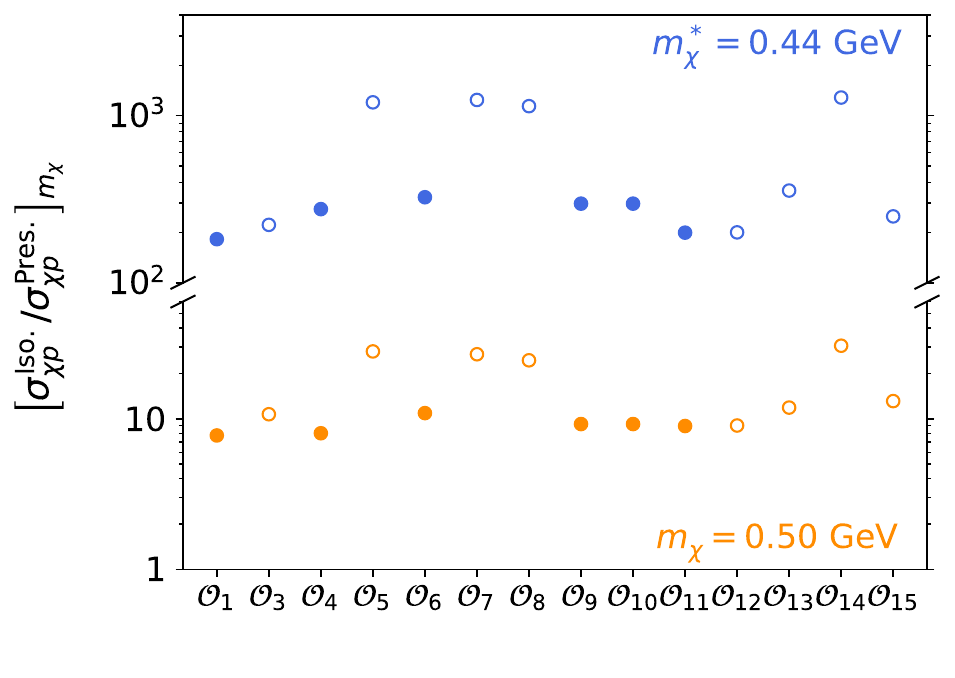}
    \end{subfigure}
    \caption{Same as figure~\ref{fig:darkside}, but for the SuperCDMS experiment.}
\label{fig:supercdms}
\end{figure*}

\begin{figure*}[h!]
    \centering
    \begin{subfigure}
        \centering
        \includegraphics[width=7.35cm]{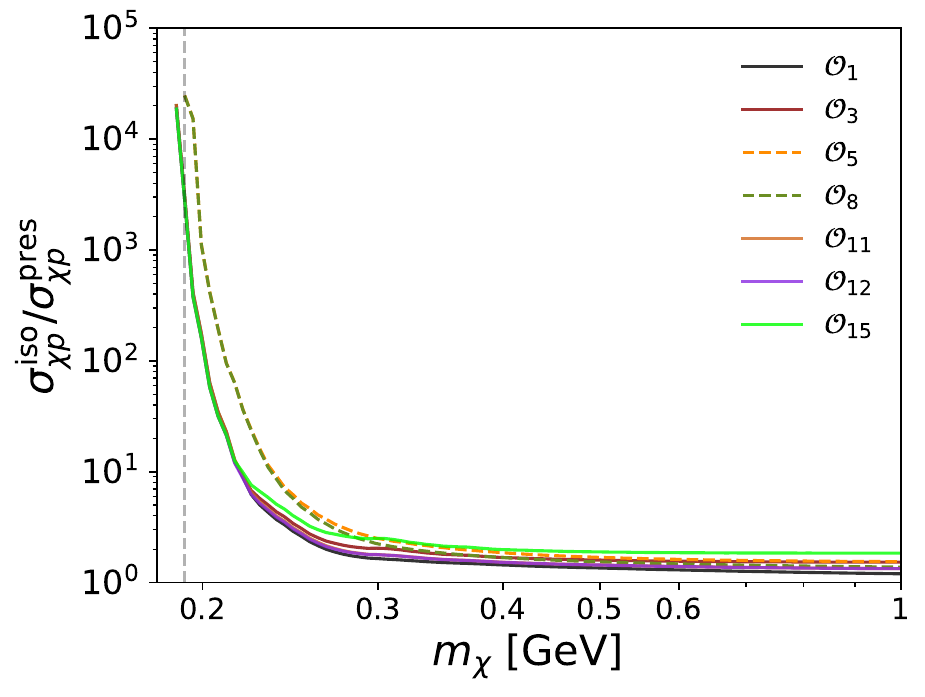}
    \end{subfigure}%
    ~ 
    \begin{subfigure}
        \centering
        \includegraphics[width=7.6cm]{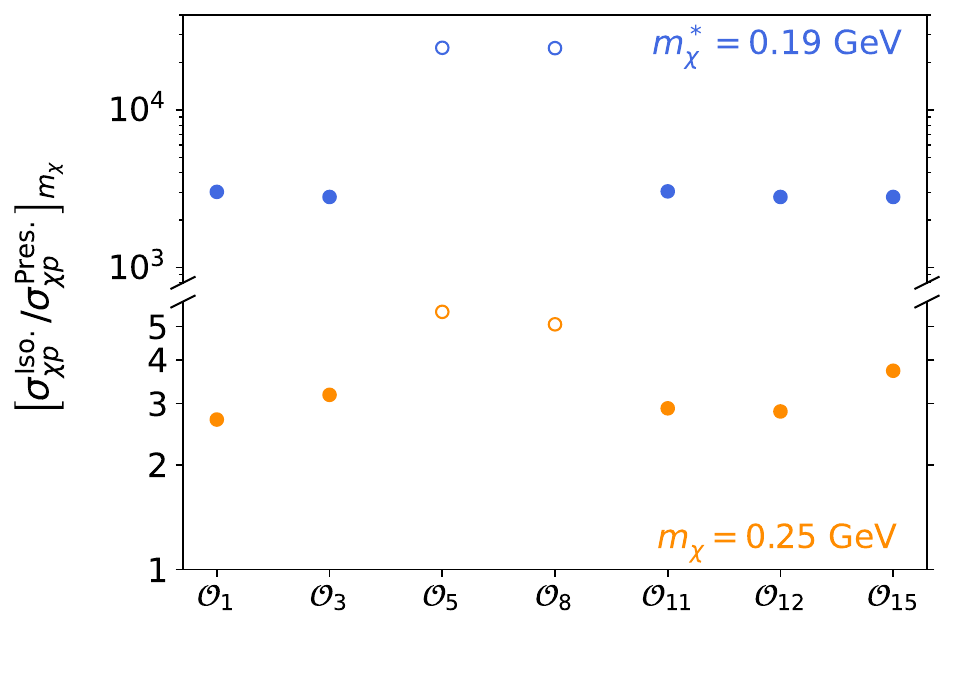}
    \end{subfigure}
    \caption{Same as figure~\ref{fig:darkside}, but for the NEWS-G experiment. }
\label{fig:newsg}
\end{figure*}

\begin{figure*}[h!]
    \centering
    \begin{subfigure}
        \centering
        \includegraphics[width=7.45cm]{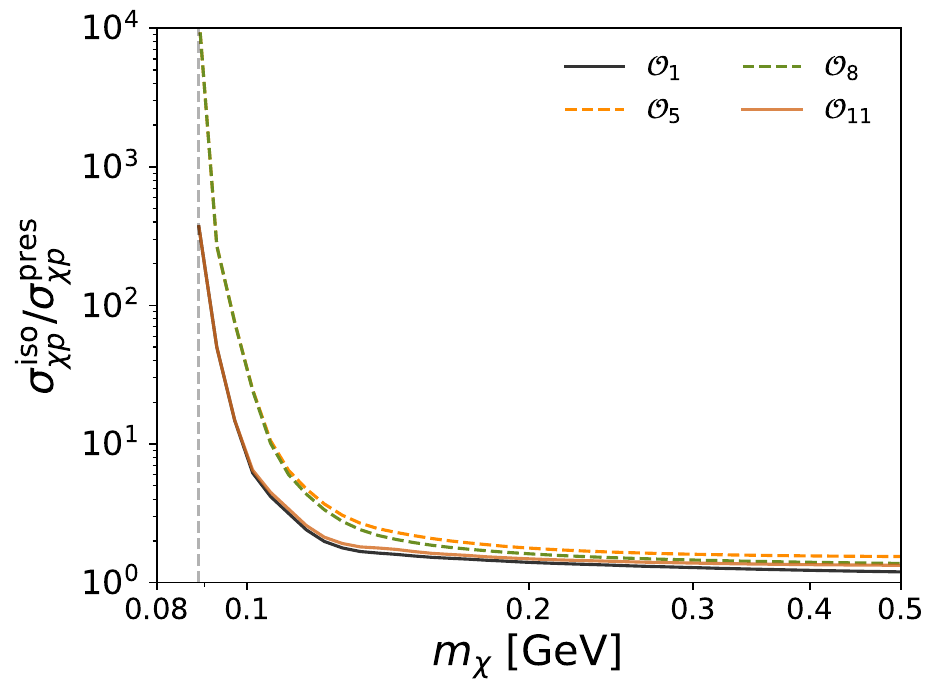}
    \end{subfigure}%
    ~ 
    \begin{subfigure}
        \centering
        \includegraphics[width=7.6cm]{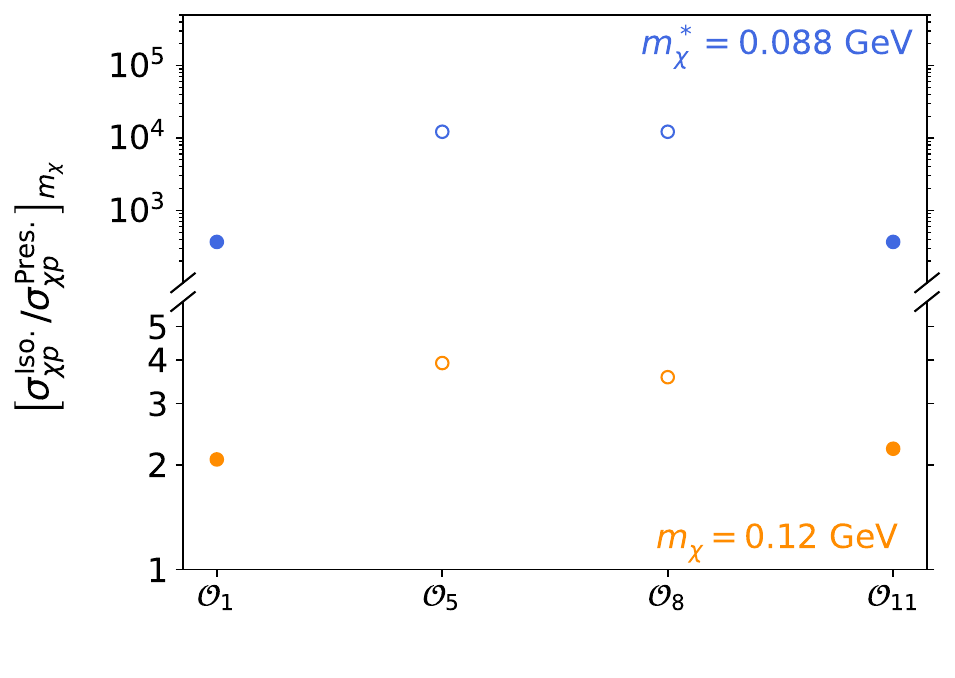}

    \end{subfigure}
    \caption{Same as figure~\ref{fig:darkside}, but for the DarkSPHERE experiment.}
\label{fig:darksphere}
\end{figure*}

 To further quantify the impact of the LMC on the exclusion limits, in the left panels of figures~\ref{fig:darkside}-\ref{fig:darksphere},  we present the ratio of 
 the upper limit at the 90\% CL on the effective DM-nucleon cross section for the isolated halo, $\sigma_{\chi p}^{\rm{iso}}$, and for the present day halo, $\sigma_{\chi p}^{\rm{pres}}$, for different operators for DarkSide-20k, SBC, DARWIN$\slash$XLZD, SuperCDMS, NEWS-G, and DarkSPHERE, respectively. Each ratio curve starts at the lowest DM mass where the ratio is finite. Below that mass, the recoil energy produced from the DM scattering in the detector is below the energy threshold of the experiment. From these plots, we can therefore quantify how the LMC impacts the exclusion limits for each operator and for the range of DM masses probed by the experiments. The right panels of figures~\ref{fig:darkside}-\ref{fig:darksphere} show the same ratio, $\sigma_{\chi p}^{\rm{iso}}/\sigma_{\chi p}^{\rm{pres}}$, evaluated at two fixed DM masses. The blue circles show the ratios evaluated at the minimum DM mass, $m_{\chi}^\ast$, for which $\sigma_{\chi p}^{\rm{iso}}$ is finite for \emph{all} operator. The value of ${m_\chi^\ast}$ for each experiment is specified with a dashed gray line in the left panels of the figures and given as a legend on the top right corner of the right panels. The orange circles show the ratios evaluated at a slightly larger but arbitrary DM mass, specified on the bottom right corner of the plots in the right panels. The ratio $[\sigma_{\chi p}^{\rm{iso}}/\sigma_{\chi p}^{\rm{pres}}]_{m_\chi}$, therefore, allows us to compare the impact of the LMC on the exclusion limits for different operators, at two fixed small DM masses close to the minimum mass that can be probed by each experiment for each operator.

 The dashed colored lines in the left panels and the open circles in the right panels of figures~\ref{fig:darkside}-\ref{fig:darksphere} identify operators for which the DM-nucleus cross section has a velocity dependent scaling compared to $\mathcal{O}_1$, as listed in the second column of table~\ref{tab:operators}. Notice that for Ar, Ne, and $^{4}{\rm He}$ which do not have spin, only operators $\mathcal{O}_5$ and $\mathcal{O}_8$ lead to a velocity-dependent scaling of the cross section. For  $\mathcal{O}_3$, $\mathcal{O}_{12}$, and $\mathcal{O}_{15}$ in case of Ar and Ne, only the $q$-dependent terms are present in this scaling. 
 
At a fixed DM mass close to the minimum mass probed by each experiment, the impact of the LMC tends to be larger for operators that lead to a velocity-dependent scaling in the DM-nucleus cross section. This is clearly the case for $\mathcal{O}_5$ and $\mathcal{O}_8$ for all six experiments. As shown in figures~\ref{fig:darkside}-\ref{fig:darksphere}, $[\sigma_{\chi p}^{\rm{iso}}/\sigma_{\chi p}^{\rm{pres}}]_{m_\chi^\ast}$ reaches $\sim 3 \times 10^5$ at $m_\chi^\ast=9.4$~GeV for DarkSide-20k, $\sim 10^7$ at $m_\chi^\ast=0.49$~GeV for SBC, $\sim 4 \times 10^9$ at $m_\chi^\ast=4.1$~GeV for DARWIN$\slash$XLZD, $\sim 10^3$ at $m_\chi^\ast=0.44$~GeV for SuperCDMS, $\sim 3 \times 10^4$ at $m_\chi^\ast=0.19$~GeV for NEWS-G, and $10^4$ at $m_\chi^\ast=0.088$~GeV for DarkSPHERE. For DARWIN$\slash$XLZD and SuperCDMS which are sensitive to all operators, the LMC's impact is also large for operators $\mathcal{O}_7$ and $\mathcal{O}_{14}$ (in addition to $\mathcal{O}_5$ and $\mathcal{O}_8$) across a range of DM masses up to $\sim 15$~GeV in DARWIN$\slash$XLZD and $\sim 1$~GeV in SuperCDMS. Notice that the value of the ratio $\sigma_{\chi p}^{\rm{iso}}/\sigma_{\chi p}^{\rm{pres}}$ is  sensitive to the DM mass at which it is evaluated. The ordering of this ratio for different operators, shown in the right panels of figures~\ref{fig:darkside}-\ref{fig:darksphere}, is especially impacted by the exact value of ${m_\chi}$.

\begin{figure*}[t]
    \centering
        \includegraphics[width=13cm]{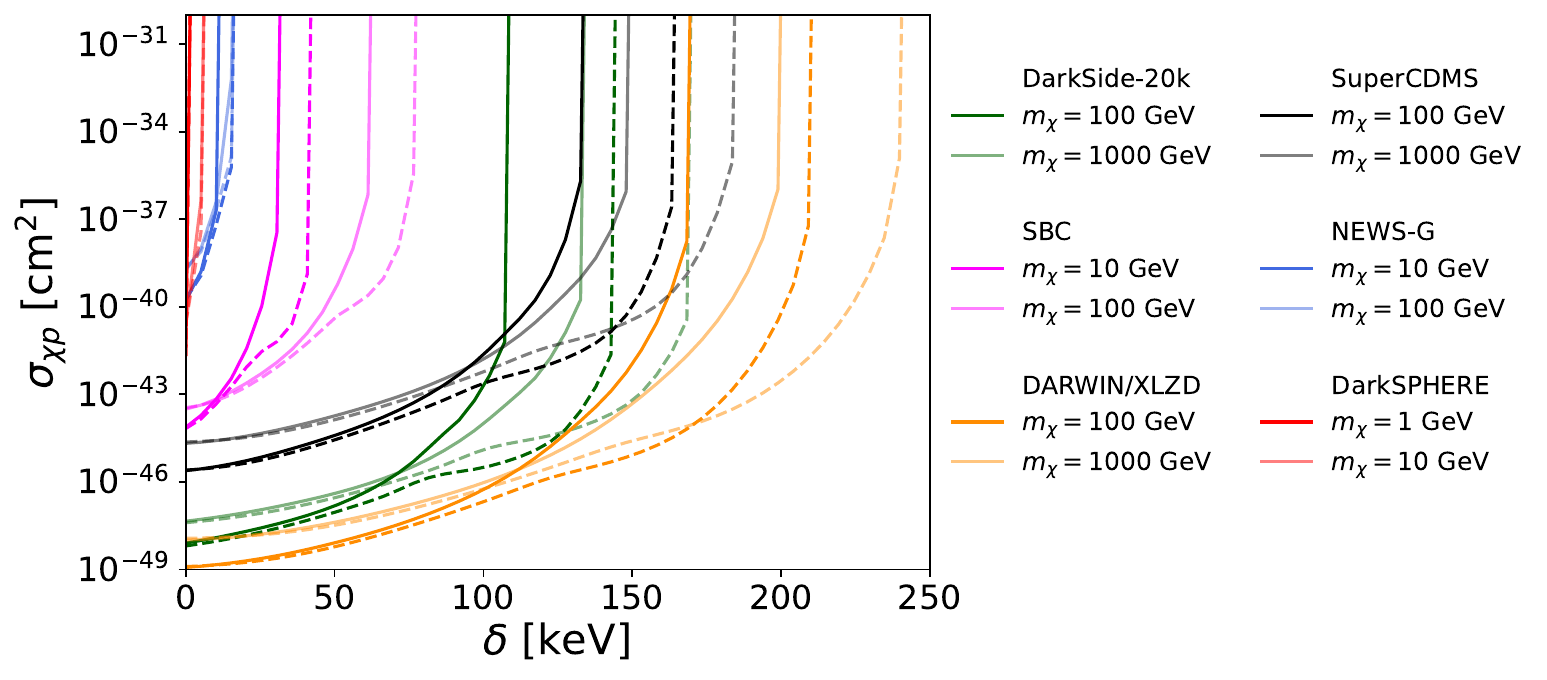} 
    \caption{Exclusion limits at the 90\% CL in the plane of the mass splitting, $\delta$, for inelastic DM and the standard spin-independent DM-nucleon cross section, $\sigma_{\chi p}$,  for DarkSide-20k (green lines), SBC (magenta lines), DARWIN$\slash$XLZD (orange lines), SuperCDMS (black lines), NEWS-G (blue lines), and DarkSPHERE (red lines). The darker (lighter) colored lines show the limits for a 100 GeV (1000 GeV) DM mass for DarkSide-20k,  DARWIN$\slash$XLZD and SuperCDMS, a 10 GeV (100 GeV) DM mass for SBC and NEWS-G, and a 1 GeV (10 GeV) DM mass for DarkSPHERE. Solid colored lines are computed using the local DM velocity distribution of the isolated MW, while dashed colored lines are computed for the present day MW-LMC analogue. The local DM density is set to $\rho_\chi = 0.3$~GeV/cm$^3$.}
\label{fig:inelastic}
\end{figure*}

Besides the case of non-standard DM interactions using NREFT, we have also studied the impact of the LMC on the standard spin-independent DM-nucleon cross section for the case of inelastic DM. In this case, the minimum DM speed, $v_{\rm min}$ (given in eq.~\eqref{eq:vmin_inel}) depends on the mass splitting, $\delta$, between the lighter and heavier DM states. To quantify the effect of the LMC on the cross section for a range of values of  $\delta$, we first fix the DM mass, taking two representative values for each experiment. Figure~\ref{fig:inelastic} shows the exclusion limits at the 90\% CL in the plane of $\delta$ and the standard spin-independent DM-nucleon cross section, $\sigma_{\chi p}$, for DarkSide-20k (green lines), SBC (magenta lines), DARWIN$\slash$XLZD (orange lines), SuperCDMS (black lines), NEWS-G (blue lines), and DarkSPHERE (red lines). For each experiment, the darker colored lines show the results for a smaller DM mass (1, 10, or 100 GeV depending on the experiment), and the lighter colored lines show the results for a larger DM mass (10, 100, or 1000 GeV depending on the experiment). The solid lines show the limits for the isolated MW analogue, while the dashed lines show the limits for the present day MW-LMC analogue. 

For all six experiments and choices of DM mass considered in figure~\ref{fig:inelastic}, the LMC shifts the exclusion limits towards larger mass splittings, $\delta$, and smaller cross sections, $\sigma_{\chi p}$. The LMC, therefore, increases the sensitivity of the experiments for probing larger values of $\delta$ in the case of inelastic DM. For each experiment, the impact of the LMC is more significant at larger $\delta$ values, where the experiment probes higher $v_{\rm min}$ and the sensitivity to the high speed tail of the halo integrals increases. Additionally, the LMC has a larger impact for experiments that have a more massive target nucleus, since they can probe a larger range of DM velocities and are sensitive to larger $\delta$ values. We can see from figure~\ref{fig:inelastic} that at a fixed large cross section, e.g.~$\sigma_{\chi p}=10^{-34}$~cm$^2$, and for both $m_\chi=100$ and 1000 GeV, the LMC shifts the exclusion limits by $\sim 40$~keV towards larger $\delta$ values for DARWIN$\slash$XLZD, which has the most massive target nucleus (i.e.~Xe) among the six experiments considered. For SuperCDMS and DarkSide, the exclusion limits shift by at most $\sim 35$~keV for the masses considered. For SBC, NEWS-G, and DarkSPHERE, the shifts are at most $\sim 15$~keV, $\sim 5$~keV, and $\sim 0.15$~keV, respectively, for the masses considered.

\begin{figure}[t]
            \includegraphics[width=.5\linewidth]{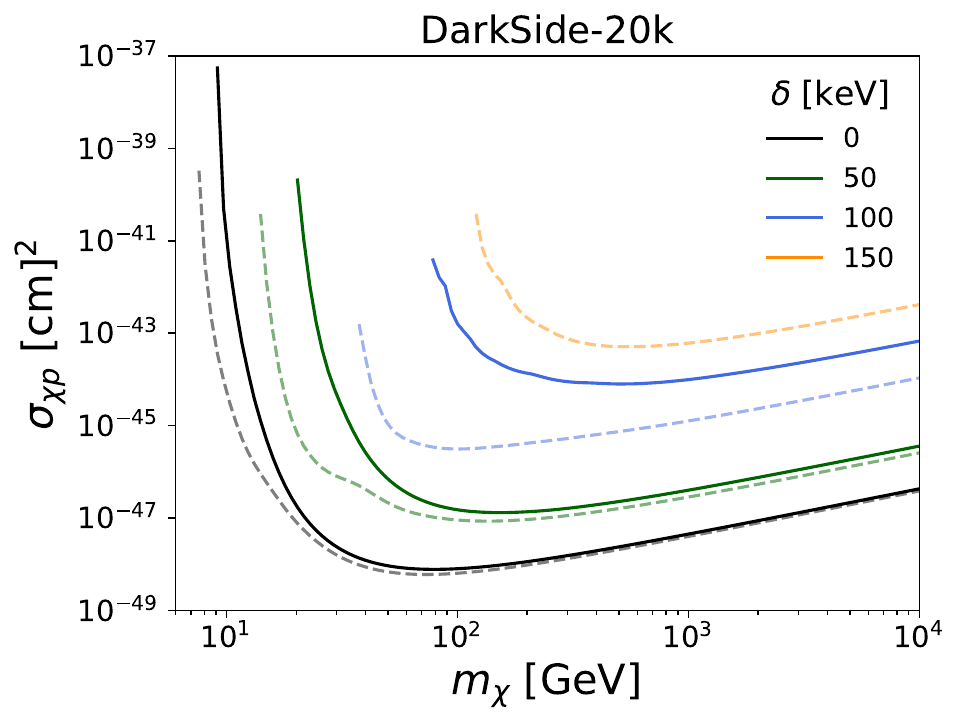}%
        \hfill
            \includegraphics[width=.5\linewidth]{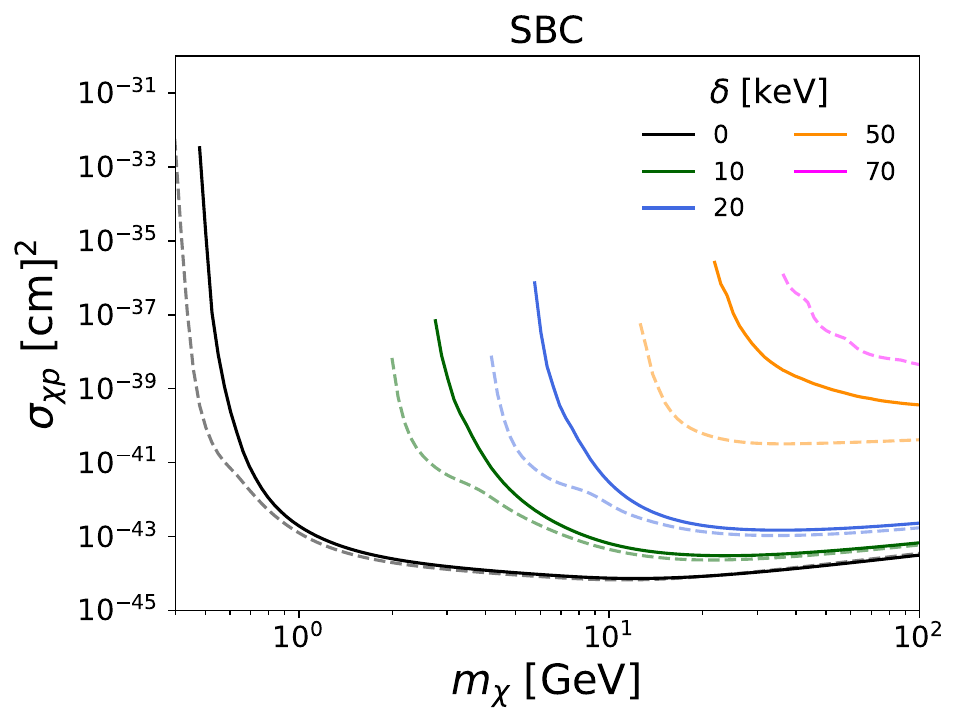}
        \\
            \includegraphics[width=.5\linewidth]{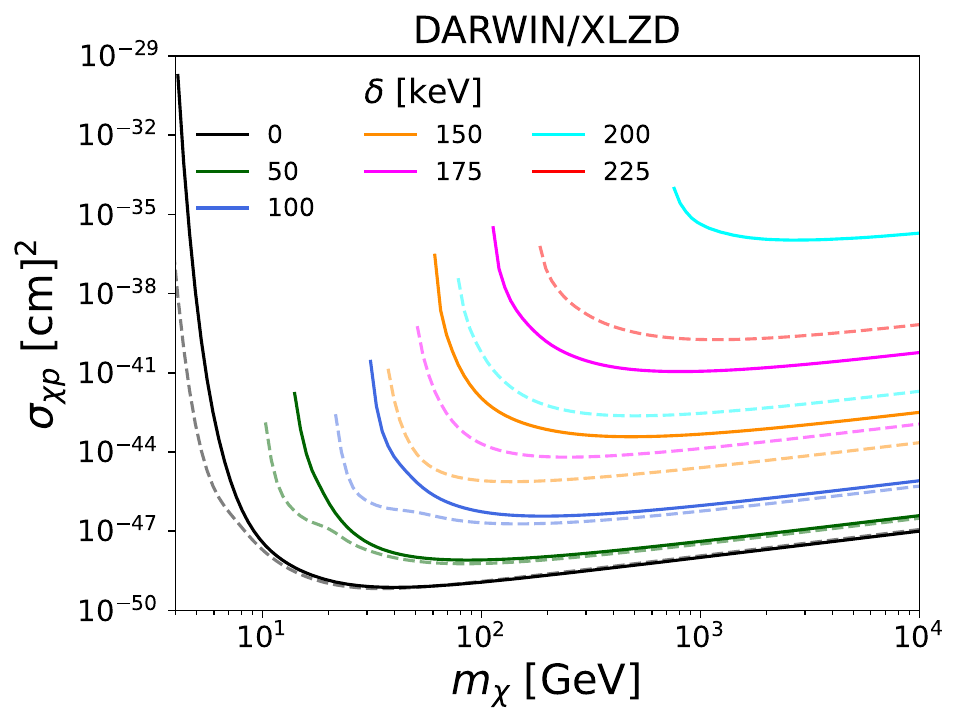}%
        \hfill
            \includegraphics[width=.5\linewidth]{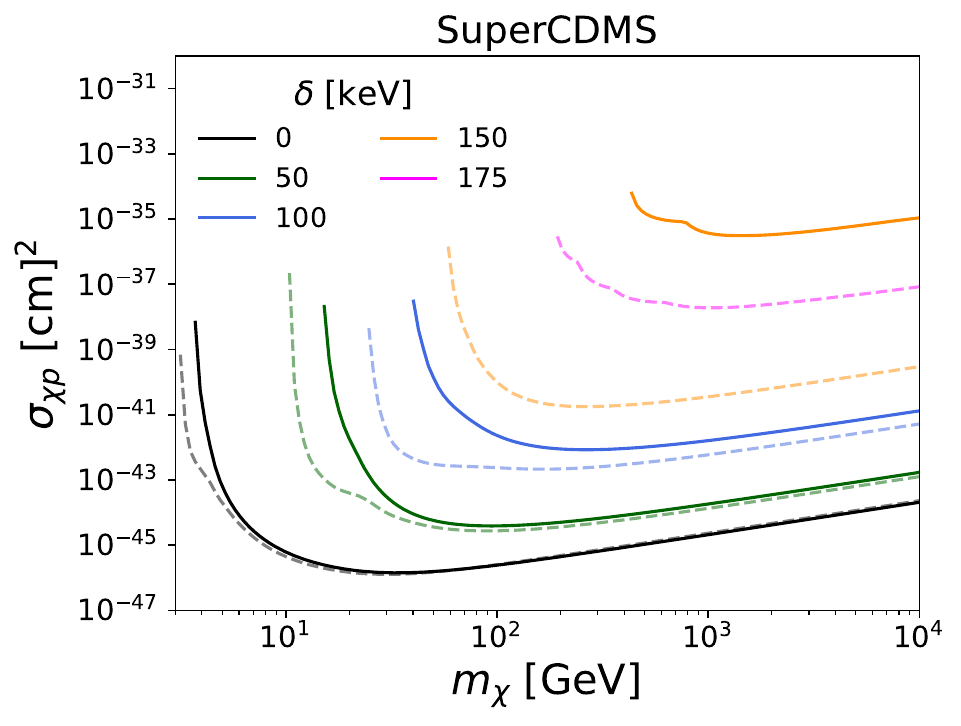}
        \\
            \includegraphics[width=.5\linewidth]{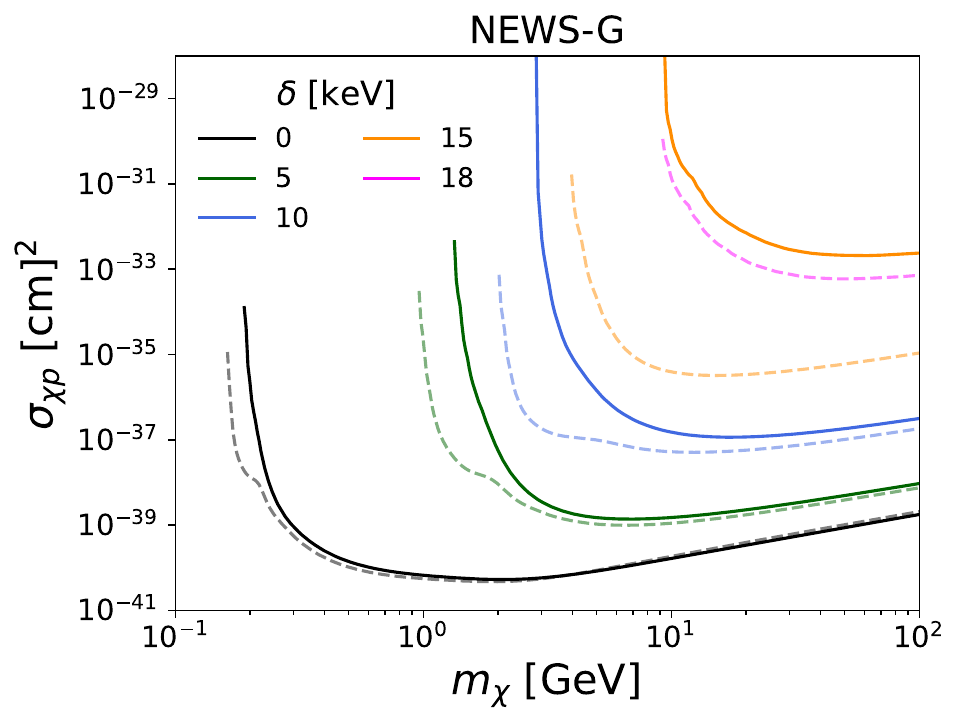}
        \hfill
            \includegraphics[width=.5\linewidth]{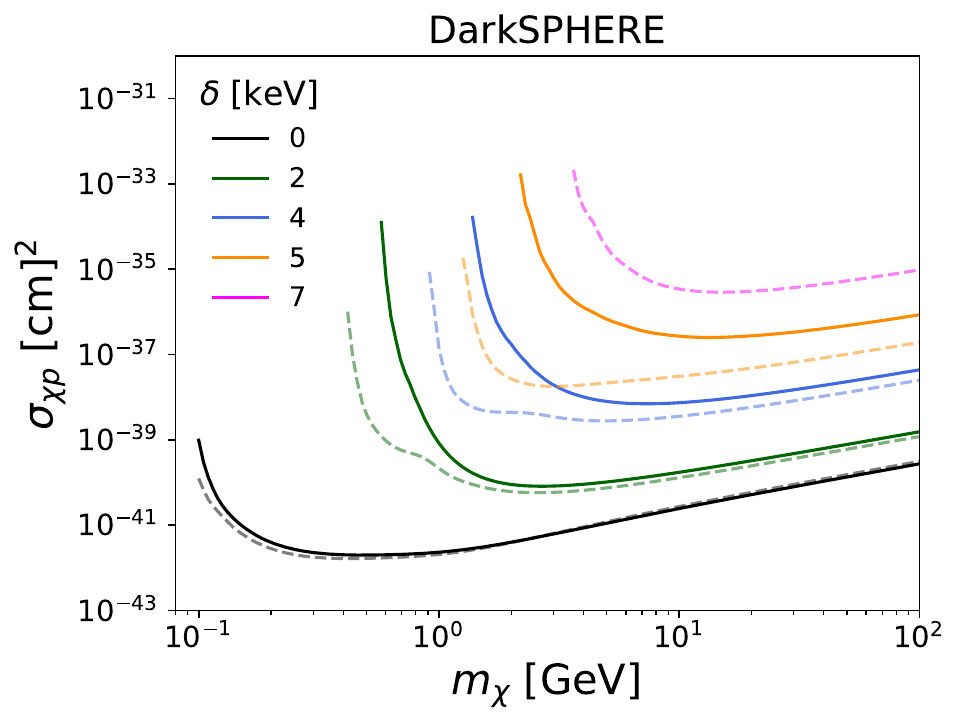}
        \caption{Exclusion limits at the 90\% CL in the plane of the DM mass, $m_\chi$, and the standard spin-independent DM-nucleon cross section, $\sigma_{\chi p}$, for DarkSide-20k (top left panel), SBC (top right panel), DARWIN$\slash$XLZD (middle left panel), SuperCDMS (middle right panel), NEWS-G (bottom left panel), and DarkSPHERE (bottom right panel). Different colored lines correspond to different values of the DM mass splitting, $\delta$, as specified in the legends. Solid colored lines are computed using the local DM velocity distribution of the isolated MW, while dashed colored lines are computed for the present day MW-LMC analogue. The local DM density is set to $\rho_\chi = 0.3$~GeV/cm$^3$. }
        \label{fig:limits_delta_all}
\end{figure}

Figure~\ref{fig:limits_delta_all} shows the exclusion limits at the 90\% CL in the $m_\chi$ and  $\sigma_{\chi p}$ plane for different values of the mass splitting $\delta$ for DarkSide-20k (top left panel), SBC (top right panel), DARWIN$\slash$XLZD (middle left panel), SuperCDMS (middle right panel), NEWS-G (bottom left panel), and DarkSPHERE (bottom right panel). The solid lines show the limits for the isolated MW, while the dashed lines show the limits for the present day MW-LMC. Similar to our results for the case of the NREFT operators, the LMC causes a shift in the exclusion limits towards smaller DM masses and smaller cross sections for inelastic DM. The effect is more prominent for larger values of $\delta$, as expected. 

We can see again from figure~\ref{fig:limits_delta_all} that the LMC extends the parameter space that can be probed by the experiments for inelastic DM. For instance, for the isolated MW, the mass splitting of $\delta=150$~keV is not accessible to DarkSide-20k. However, when the impact of the LMC is included (i.e.~for the present day MW-LMC analogue), DarkSide-20k can probe this mass splitting. Similarly, mass splittings of $\delta=70$, 225, 175, 18, and 7~keV, become accessible to the SBC, DARWIN$\slash$XLZD,  SuperCDMS, NEWS-G, and DarkSPHERE experiments, respectively, only when the impact of the LMC is considered.

\section{Conclusions}
\label{sec:conclusions}

In this work we have studied the impact of the LMC on the interpretation of results from near-future direct detection experiments for non-standard DM interactions, using a simulated MW-LMC system from the Auriga magneto-hydrodynamical simulations. In particular, we investigated how the LMC affects the predicted signals in DarkSide-20k,  SBC, DARWIN$\slash$XLZD, SuperCDMS, NEWS-G, and DarkSPHERE, considering DM-nucleon effective interactions using the NREFT approach, as well as inelastic DM scattering. These experiments use different detector technologies and target nuclei, and as a result have different sensitivities to various NREFT operators, as well as the impact of the LMC. 

We utilized the simulated MW-LMC system first studied in ref.~\cite{Smith-Orlik:2023kyl} and compared our results for two simulation snapshots: the isolated MW analogue (before the infall of the LMC into the MW), and the present day MW-LMC analogue. As shown in previous work employing both cosmological and idealized simulations~\cite{Smith-Orlik:2023kyl, Besla:2019xbx, Donaldson:2021byu}, the LMC significantly boosts the high speed tail of the local DM velocity distribution.  This leads to a shift of $\sim 150$~km/s in the tails of the halo integrals, $\eta$ (eq.~\eqref{eq:etavmin}) and $h$ (eq.~\eqref{eq:hvmin}), towards higher speeds. These two halo integrals encode the dependence of the direct detection event rate on the local DM velocity distribution for general non-standard DM interactions. 

The shifts in the high speed tails of the halo integrals due to the influence of the LMC cause significant shifts in direct detection exclusion limits towards smaller effective DM-nucleon cross sections and DM masses for all NREFT operators and the six experiments considered. As expected, the impact of the LMC tends to be more pronounced for operators that lead to a velocity-dependent scaling in the DM-nucleus cross section. For the case of the DarkSide-20k, SBC, NEWS-G, and DarkSPHERE experiments which employ spin zero Ar, Ne, and $^4{\rm He}$ targets, the greatest shifts in the exclusion limits occur for the two velocity-dependent operators present, $\mathcal{O}_5$ and $\mathcal{O}_8$. In particular, the exclusion limits shift by more than 5 orders of magnitude in DarkSide-20k at a DM mass of $\sim 9$~GeV, 7 orders of magnitude in SBC at $\sim 0.5$~GeV, more than 4 orders of magnitude in NEWS-G at $\sim 0.2$~GeV, and 4 orders of magnitude in DarkSPHERE at $\sim 0.09$~GeV.

For the case of DARWIN$\slash$XLZD and SuperCDMS, the largest impact on the exclusion limits happens for  the velocity-dependent $\mathcal{O}_5$, $\mathcal{O}_7$, $\mathcal{O}_8$, and $\mathcal{O}_{14}$ operators across a range of DM masses up to $\sim 15$~GeV for DARWIN$\slash$XLZD and up to $\sim 1$~GeV for SuperCDMS. At the lowest DM mass that can be probed by all operators, the largest shifts in the exclusion limits occur for $\mathcal{O}_5$ and $\mathcal{O}_8$ in DARWIN$\slash$XLZD, where the limits shift by more than 9 orders of magnitude at a DM mass of $\sim 4$~GeV. For SuperCDMS, operators $\mathcal{O}_5$, $\mathcal{O}_7$, $\mathcal{O}_8$, and $\mathcal{O}_{14}$, show a similar shift of 3 orders of magnitude at $\sim 0.4$~GeV. In both experiments, the remaining velocity-dependent operators, $\mathcal{O}_{3}$, $\mathcal{O}_{12}$, $\mathcal{O}_{13}$, and $\mathcal{O}_{15}$, show a smaller but still significant shift in the cross sections. The shift for this latter group of velocity-dependent operators becomes comparable to that for a number of velocity-independent operators, such as $\mathcal{O}_{6}$, which has a $q^4$ scaling in the cross section.

We also studied the impact of the LMC on direct detection of inelastic DM scattering, where the DM particle up-scatters to a heavier state with a mass splitting $\delta$. At a fixed DM mass, the LMC significantly shifts the direct detection exclusion limits  towards smaller cross sections and larger $\delta$, with the impact being more pronounced for experiments with a more massive target nucleus. Considering a fixed $\delta$ value, the exclusion limits are shifted towards smaller DM masses and smaller cross sections by several orders of magnitude, due to the impact of the LMC. The LMC, therefore, allows the experiments to probe higher values of $\delta$. In particular, because of LMC's impact, DARWIN$\slash$XLZD, SuperCDMS, DarkSide, SBC, NEWS-G, and DarkSPHERE would be able to probe $\delta$ values as large as 225, 175, 150, 70, 18, and 7 keV, respectively. 

Our results demonstrate that the presence of the LMC significantly expands the low-mass parameter space that can be probed by direct detection experiments towards smaller DM-nucleon cross sections, for all NREFT operators. Furthermore, the LMC greatly enhances the experiments' sensitivity to larger values of the mass splitting for inelastic DM.

\acknowledgments

JR and NB acknowledge the support of the Natural Sciences and Engineering Research Council of Canada (NSERC), funding reference number RGPIN-2020-07138, and the NSERC Discovery Launch Supplement, DGECR-2020-00231. NB acknowledges the support of the Canada Research Chairs Program. We thank Azadeh Fattahi for providing the re-simulated Auriga halo studied in this work. We thank Daniel Durnford for fruitful discussions regarding the cross-sections for NEWS-G and DarkSPHERE experiments. We also thank Knut Dundas Morå for helpful discussion regrading DARWIN/XLZD cross-section limits. We thank, as well, Simon Viel, Shawn Westerdale and Ariel Zuñiga Reyes for helpful discussion on the NREFT operators obtained for the DEAP experiment.

\appendix

\typeout{}
\bibliographystyle{JHEP}
\bibliography{refs}

\end{document}